\documentclass{article}

\usepackage{graphicx}      
\usepackage{amsmath}	   
\usepackage{amsthm}
\usepackage{amssymb}
\usepackage{algorithm}
\usepackage{enumerate}
\usepackage{xcolor}
\usepackage{pgfplots}
\usepackage{setspace}
\usepackage{xspace}
\pgfplotsset{compat=newest} 
\pgfplotsset{plot coordinates/math parser=false} 
\newlength\figureheight 
\newlength\figurewidth

\pgfplotsset{grid style={dotted,black!50!white}}

\usepackage{float}
\usepackage{subfig}
\usepackage[a4paper,bottom=2.54cm,top=2.54cm]{geometry}
\usepackage{parskip}
\newcommand{\coverAuthors}{Jack Umenberger, Johan W{\aa}gberg, Ian R. Manchester and Thomas B. Sch{\"o}n}

\newtheorem{thm}{Theorem}
\newtheorem{lem}[thm]{Lemma}
\newtheorem{prop}[thm]{Proposition}
\newtheorem{rmk}{Remark}
\newenvironment{pf}{\begin{proof}}{\end{proof}}

\newcommand{\ie}{i.e.\xspace}
\newcommand{\myd}{\mathrm{d}}

\newcommand{\myth}{^\textup{th}}
\newcommand{\real}{\mathbb{R}}

\newcommand{\tha}{\alpha}
\newcommand{\thb}{\beta}
\newcommand{\thc}{\gamma}

\newcommand{\lagj}{\bar{J}}
\newcommand{\qbnd}{\bar{Q}_3}
\newcommand{\loglik}{L_\theta(y_{1:T})}

\newcommand{\qls}{Q_{\textup{ls}}}
\newcommand{\qld}{Q_{\textup{ld}}}
\newcommand{\ththk}{(\theta,\theta_k)}

\newcommand{\mult}{\lambda}
\newcommand{\searg}{\gamma}

\newcommand{\expsym}{\operatorname{E}}

\newcommand{\Expb}[2]{\expsym_{#1}\left[#2\right]}
\newcommand{\CExpb}[3]{\Expb{#1}{#2\,\middle\vert\,#3}}

\newcommand{\varsym}{\operatorname{Var}}

\newcommand{\Varb}[2]{\varsym_{#1}\left[#2\right]}
\newcommand{\CVarb}[3]{\Varb{#1}{#2\,\middle\vert\,#3}}

\newcommand{\initstate}{\xi_1}

\begin{document}
\pagenumbering{gobble}
\begin{titlepage}

\title{Linear System Identification via EM with Latent Disturbances and Lagrangian Relaxation}

\author{
	Jack Umenberger\thanks{School of Aerospace, Mechanical and Mechatronic Engineering, University of Sydney, Australia e-mail: \{j.umenberger, i.manchester\}@acfr.usyd.edu.au} 
	\and Johan W{\aa}gberg\thanks{Department of Information Technology, Uppsala University, Sweden, e-mail: \{johan.wagberg, thomas.schon\}@it.uu.se.}
	\and Ian R. Manchester\footnotemark[1]
	\and Thomas B. Sch\"on\footnotemark[2]}

\makeatletter
\let\Title\@title
\let\Author\@author
\makeatother

\maketitle

\begin{center}
\fbox{
\begin{minipage}{.95\columnwidth}
\renewcommand\labelitemi{\color{red}\large$\bullet$}
\begin{itemize}
\item {\large \textbf{Please cite this version:}} \\[0.2cm]
	\coverAuthors. \textbf{\Title.} \textit{arXiv, 2016}.
\end{itemize}
\end{minipage}}
\end{center}

\begin{abstract}               
In the application of the Expectation Maximization algorithm to identification of dynamical systems, internal \emph{states} are typically chosen as latent variables, for simplicity. In this work, we propose a different choice of latent variables, namely, system \emph{disturbances}. Such a formulation elegantly handles the problematic case of singular state space models, and is shown, under certain circumstances, to improve the fidelity of bounds on the likelihood, leading to convergence in fewer iterations. To access these benefits we develop a Lagrangian relaxation of the nonconvex optimization problems that arise in the latent disturbances formulation, and proceed via semidefinite programming.  
\end{abstract}

\end{titlepage}

\pagenumbering{arabic}
\section{Introduction}
Linear time invariant (LTI) state-space models provide a useful approximation of dynamical system behavior in a
multitude of applications. In situations where models cannot be derived from first principles, some form of data-driven modeling, i.e. system identification, is appropriate \cite{Ljung1999}. This paper is concerned with identification of discrete-time LTI models of the form
\begin{subequations}\label{eq:lgss}
  \begin{align} 
    x_{t+1} &=Ax_t +Bu_t + Gw_t, \label{eq:lgss_dynamics} \\
    y_t &= Cx_t +Du_t+v_t, \label{eq:lgss_output}
  \end{align}
\end{subequations}
where $x_t\in\mathbb{R}^{n_x}$ denotes the system state, and $u_t\in\mathbb{R}^{n_u}$, $y_t\in\mathbb{R}^{n_y}$ denote the observed input and output, respectively. The disturbances (a.k.a. process noise), $w_t\in\real^{n_w}$ and measurement noise, $v_t$, are modeled as zero mean Gaussian white noise processes, while the uncertainty in the initial condition $x_1$ is modeled by a normal distribution, i.e.
\begin{equation*}
w_t\sim\mathcal{N}(0,\Sigma_w), \ 
v_t\sim\mathcal{N}(0,\Sigma_v), \
x_1\sim\mathcal{N}(\mu,\Sigma_1).
\end{equation*}
For convenience, all unknown model parameters are denoted by the variable $\theta=\lbrace \mu,\Sigma_1,\Sigma_w,\Sigma_v, $\\$A,B,G,C,D\rbrace$.

\par Despite the simplicity of LTI models, identification of such systems is complicated by the presence of \emph{latent} variables. Specifically, in applications the observed data typically consists of inputs and (noisy) outputs, but not internal states or exogenous disturbances. 

\par Various strategies have been developed to deal with this `missing data'. \emph{Marginalization}, for instance, involves integrating out (i.e. marginalizing over) the latent variables, leaving $\theta$ as the only quantity to be estimated. This is the approach adopted by prediction error methods \cite{Ljung1999,ljung2002prediction} and the Metropolis-Hastings algorithm \cite{metropolis1953equation,hastings1970monte}. 

\par Alternatively, one may treat the latent variables as additional quantities to be estimated together with the model parameters. Such a strategy is termed \emph{data augmentation}, and examples include subspace methods \cite{Vandenberghet1996,larimore1983system}, the Gibbs sampler \cite{geman1984stochastic}, and the {Expectation Maximization} (EM) algorithm \cite{dempster1977maximum,Schon2011,gibson2005robust,shumway1982approach}.

\par Recently, a new family of methods have been developed in which one \emph{supremizes over} the latent variables to obtain convex upper bounds for quality-of-fit cost functions, such as simulation error (a.k.a. output error) \cite{Tobenkin2010,Bond2010,Megretski2008}. An important technique employed in this approach is Lagrangian relaxation \cite{lemarechal2001lagrangian,yakubovich1971s,polik2007survey}, which replaces difficult constrained optimization problems with tractable, unconstrained, convex approximations.      

\par This work draws on the underlying similarities between EM and Lagrangian relaxation to develop a new algorithm that seeks the maximum likelihood estimate of the model parameters $\theta$, given measurements $u_{1:T}$ and $y_{1:T}$, i.e.
\begin{equation}\label{eq:maxLikeProb}
\theta^{\text{ML}}=\arg\max_{\theta}p_{\theta}(u_{1:T},y_{1:T}).
\end{equation}

\par The EM algorithm is an iterative approach to ML estimation, in which estimates of the latent variables are used to construct tractable lower bounds to the likelihood. In the application of EM to (\ref{eq:maxLikeProb}) the latent variables are typically taken to be the system \emph{states}, $x_{1:T}$, as this simplifies the ensuing optimization problem(s). Specifically, optimization of the bound at each iteration reduces to linear least squares \cite{gibson2005robust}.

\par In this work, we formulate the EM algorithm, for the approximate solution of (\ref{eq:maxLikeProb}), over latent \emph{disturbances}, $w_{1:T}$. In contrast to the latent states formulation, the optimization of bounds based on disturbances is nonconvex. By applying Lagrangian relaxation, we obtain new bounds that can be optimized by semidefinite programming (SDP). The resulting algorithm can be considered an example of the more general \emph{minorization maximization} principle \cite{ortega1970iterative,hunter2004tutorial}.

\par The rewards for this additional complexity are threefold. First, the proposed method elegantly handles identification of \emph{singular} state-space models (i.e. $n_w<n_x$), a case to which the standard formulation of EM over latent states is not applicable, without modification. Secondly, this approach naturally ensures stability of the model at each iteration. Finally, when the \emph{magnitude} of the disturbances (i.e. $\Sigma_w$) is small, we show that use of latent disturbances produces better approximations to the likelihood, leading to convergence in fewer iterations.

\par We first introduced the basic idea of combining Lagrangian relaxation with a formulation of EM over latent disturbances in our conference paper \cite{umenberger2015em}. This paper extends this recent work in several significant ways. In Section \ref{sec:lr_mstep} we apply Lagrangian relaxation without resorting to Monte Carlo approximations, unlike the approach outlined in \cite{umenberger2015em}. Furthermore, the Lagrangian relaxation detailed in this paper makes use of a more sophisticated multiplier, introduced in Section \ref{sec:multipliers}. A new study of the behavior of the EM algorithm for large and small disturbances is presented in Sections \ref{sec:limit_cases} and \ref{sec:1dbounds}, offering insights into the results of numerical experiments on convergence rates in Section \ref{sec:conv_rates}. 

\section{Preliminaries}

\subsection{Notation}
The cone of real, symmetric nonnegative (positive) definite matrices is denoted by $\mathbb{S}^{n}_{+}$ ($\mathbb{S}^{n}_{++}$). The $n\times n$ identity matrix is denoted $I_n$. Let $\text{vec}:\mathbb{R}^{m\times n}\mapsto\mathbb{R}^{mn}$ denote the function that stacks the columns of a matrix to produce a column vector. The Kronecker product is denoted $\otimes$. The transpose of a matrix $A$ is denoted $A'$, and $|A|_Q^2$ is shorthand for $A'QA$. Time series data $\lbrace x_t\rbrace_{t=a}^b$ is denoted $x_{a:b}$ where $a,b\in\mathbb{N}$. A random variable $x$ distributed according to the multivariate normal distribution, with mean $\mu$ and covariance $\Sigma$, is denoted $x\sim\mathcal{N}(x;\mu,\Sigma)$. We use $a(\theta)\propto b(\theta)$ to mean $b(\theta)=c_1a(\theta)+c_2$ where $c_1, c_2$ are constants that do not effect optimization of $a(\theta)$ w.r.t. $\theta$. For invertible $A$, $A\backslash B$ is shorthand for $A^{-1}B$. The log likelihood function is denoted $L_\theta(y_{1:T})\triangleq\log p_\theta(u_{1:T},y_{1:T})$.

\subsection{The minorization-maximization principle}
The minorization-maximization (MM) principle \cite{ortega1970iterative,hunter2004tutorial} is an iterative approach to optimization problems of the form $\max_\theta \ f(\theta)$. Given an objective function $f(\theta)$ (not necessarily a likelihood), at each iteration of an MM algorithm we first build a \emph{tight} lower bound $b(\theta,\theta_k)$ satisfying 
\begin{equation*}
f(\theta) \geq b(\theta,\theta_k) \ \forall \ \theta \text{ and } f(\theta_k)=b(\theta_k,\theta_k),
\end{equation*}
i.e. we \emph{minorize} $f$ by $b$. Then we optimize $b(\theta,\theta_k)$ w.r.t. $\theta$ to obtain $\theta_{k+1}$ such that $f(\theta_{k+1})\geq f(\theta_k)$. The principle is useful when direct optimization of $f$ is challenging, but optimization of $b$ is tractable (e.g. concave). In the following two subsections, we present EM and Lagrangian relaxation as special cases of the MM principle, for problems involving missing data. Each of these algorithms is predicated on the assumption that there exists latent variables, $z$, such that optimization of $f(\theta)$ would be more straightforward if $z$ were known. 

\subsection{The Expectation Maximization algorithm}\label{sec:EM}
The EM algorithm \cite{dempster1977maximum} applies the MM principle to ML estimation, \ie $f(\theta)=L_\theta(y_{1:T})$. Each iteration of the algorithm consists of two steps: the expectation (E) step computes the \emph{auxiliary} function
\begin{align}\label{eq:Q_basic}
  Q(\theta,\theta_k) &\triangleq\int L_{\theta}(y_{1:T},Z) p_{\theta_k}(Z \mid y_{1:T}) \ \myd Z  = \CExpb{\theta_k}{L_{\theta}(y_{1:T},Z)}{y_{1:T}}, 
\end{align}
which is then maximized in lieu of the likelihood function during the maximization (M) step. The auxiliary function can be shown to satisfy the following inequality
\begin{equation}\label{eq:dLgdQ}
L_\theta(y_{1:T})-L_{\theta_k}(y_{1:T}) \geq Q(\theta,\theta_k)-Q(\theta_k,\theta_k)
\end{equation}
and so the new parameter estimate $\theta_{k+1}$ obtained by maximization of $Q(\theta,\theta_k)$ is guaranteed to be of equal or greater likelihood than $\theta_k$. In this sense, EM may be thought of as a specific MM recipe for building lower bounds $Q(\theta,\theta_k)$ to the objective $L_\theta(y_{1:T})$, in ML estimation problems involving latent variables.

\begin{rmk}
Strictly speaking $Q(\theta,\theta_k)$ does not minorize $L_\theta(y_{1:T})$. Rather, the \emph{change} in $Q(\theta,\theta_k)$ lower bounds the \emph{change} in $L_\theta(y_{1:T})$; c.f. (\ref{eq:dLgdQ}). Nevertheless, with some abuse of terminology, we will refer to $Q(\theta,\theta_k)$ as a lower bound, as shorthand for the relationship in (\ref{eq:dLgdQ}).
\end{rmk} 

\subsection{Lagrangian relaxation}\label{sec:LRintro}
The technique of Lagrangian relaxation applies the MM principle to \emph{constrained} optimization problems of the form 
\begin{equation}\label{eq:basicLRproblem}
\min_{\theta,z} \ J(\theta,z) \text{ s.t. } F(\theta,z)=0,  
\end{equation} 
i.e. $f(\theta)=J(\theta,z^*)$ where $z^*$ is such that $F(\theta,z^*)=0$. Here $J(\theta,z)$ is a cost function assumed to be convex in $\theta$, and $F(\theta,z)$, assumed affine in $\theta$, encodes the constraints. Notice that we present the problem as cost minimization, rather than objective maximization, and consequently develop upper bounds; however, this difference in superficial.

\par Unlike EM, in which we estimate $z$, Lagrangian relaxation \emph{supremizes} over the latent variables to generate the bound. Specifically, the relaxation of (\ref{eq:basicLRproblem}) takes the form
\begin{equation}\label{eq:LRgeneral}
\lagj_\mult(\theta)=\sup_z \ J(\theta,z) + \lambda F(\theta,z),
\end{equation}
where $\mult$ may be interpreted as a Lagrange multiplier. For arbitrary $\mult$, the function $\lagj_{\mult}(\theta)$ has two key properties:
\begin{enumerate}[1)]
\item It is convex in $\theta$. Recall that $J$ and $F$ are convex and affine in $\theta$, respectively. As such, $\lagj_{\mult}(\theta)$ is the supremum of an infinite family of convex functions, and is, therefore, itself convex in $\theta$; see Section 3.2.3 of \cite{Boyd2004}.
\item It is an upper bound for the original problem (\ref{eq:basicLRproblem}). Given $\theta$, let $z^*$ be such that $F(\theta,z^*)=0$. Then
\begin{align*}
J(\theta,z^*) + \lambda F(\theta,z^*) = J(\theta,z^*) = f(\theta),
\end{align*}
which implies that the supremum over all $z$ can be no smaller; i.e. $\lagj_\lambda(\theta)$ is an upper bound for $f(\theta)$.
\end{enumerate} 
The original optimization problem (\ref{eq:basicLRproblem}) may then be approximated by the convex program $\min_\theta \ \lagj_\lambda(\theta)$.

\subsection{Latent variables for dynamical systems} \label{sec:ZforDynamical}
In the application of EM to the identification of dynamical systems, there are two possible choices of latent variables: systems states, $x_{1:T}$, and initial conditions and disturbances $\lbrace x_1,w_{1:T}\rbrace$. Choosing latent states yields a joint likelihood function of the form
\begin{equation}\label{eq:jointDistLVS}
p_{\theta}(y_{1:T},x_{1:T})= \bigg[ \prod_{t=1}^{T} p_{\theta}(y_t \mid  x_t) \bigg] \bigg[ \prod_{t=1}^{T-1} p_{\theta}(x_{t+1} \mid x_t) \bigg]  p_{\theta}(x_1),
\end{equation} 
whereas latent disturbances leads to  
\begin{equation}\label{eq:jointDistLVN_prod}
p_{\theta}(y_{1:T},x_1,w_{1:T}) = \bigg[ \prod_{t=1}^{T} p_{\theta}(y_t  \mid x_t) \bigg]p_{\theta}(w_{1:T})p_{\theta}(x_1),
\end{equation}
where $x_{t+1}=Ax_t+Bu_t+Gw_t$ for $t=1,\dots,T$. We denote this \emph{simulated} state sequence by 
\begin{align}
\mathcal{X}_T(\theta,u_{1:T},{x}_1,w_{1:T}) = \Big\lbrace x_{1:T}: x_t = A^{t-1}x_1+\sum_{\tau=1}^{t-1}A^{t-1-\tau}(Bu_\tau+Gw_\tau)\Big\rbrace,
\end{align}
which, for given $\theta$, is a \emph{deterministic} mapping from initial conditions and disturbances to system states.

\par One can begin to understand the relationship between the choice of latent variables and difficulty of the ensuing optimization problems by examining the joint log likelihood. For latent states, $\log p_\theta(y_{1:T},x_{1:T})$ decomposes as
\begin{align*}
  \log p_{\theta}(y_{1:T},x_{1:T})
    &= \sum_{t=1}^{T}|y_t-Cx_t-Du_t|_{\Sigma_v^{-1}}^2 + \sum_{t=1}^{T-1}|x_{t+1}-Ax_t-Bu_t|^2_{\Sigma_w^{-1}} \\ 
&\quad + |x_1-\mu|_{\Sigma_1^{-1}}^2 + T\log\det\Sigma_v \nonumber + T\log\det\Sigma_w +  \log\det\Sigma_1
\end{align*}
Given $\lbrace u_{1:T},x_{1:T},y_{1:T}\rbrace$, optimization of this function (w.r.t. $\theta$) amounts to simple linear least squares. Conversely, for latent disturbances, $\log p_\theta(y_{1:T},x_1,w_{1:T})$, is given by 
\begin{equation}
  \begin{aligned}
    \log p_{\theta}(y_{1:T},x_1,w_{1:T})&=\mathcal{E}(\theta,u_{1:T},y_{1:T},x_1,w_{1:T}) + \sum_{t=1}^{T}|w_t|_{\Sigma_w^{-1}}^2 + |x_1-\mu|_{\Sigma_1^{-1}}^2 \\ &\quad+ T\log\det\Sigma_v + T\log\det\Sigma_w +   \log\det\Sigma_1.
  \end{aligned}
  \label{eq:loglik_lvn_decomp}
\end{equation}
Here $\mathcal{E}(\theta,u_{1:T},y_{1:T},x_1,w_{1:T})$ denotes the \emph{simulation error}, defined 
\begin{equation}\label{eq:sim_error}
\mathcal{E}(\theta,u_{1:T},y_{1:T},x_1,w_{1:T})\triangleq\sum_{t=1}^{T}|y_t-Cx_t-Du_t|_{\Sigma_v^{-1}}^2
\end{equation}
where $x_{1:T}=\mathcal{X}_T(\theta,u_{1:T},x_1,w_{1:T})$.
This dependence on the simulated state sequence renders optimization of (\ref{eq:loglik_lvn_decomp}) a challenging nonlinear, nonconvex problem.

\section{EM with latent disturbances}\label{sec:EM_lvn}

In this section we detail the application of EM to the identification of LGSS models, when formulated with latent disturbances; refer to \cite{gibson2005robust} for the formulation over latent states. Each iteration of the algorithm involves optimization of the \emph{auxiliary function}
\begin{equation}\label{eq:Q_lvn}
  \begin{aligned}
  Q(\theta,\theta_k) 
    &= \CExpb{\theta_k}{\log p_\theta(y_{1:T},x_1,w_{1:T})}{y_{1:T}} \\
    &= \int \log p_\theta(y_{1:T},x_1,w_{1:T})p_{\theta_k}(x_1,w_{1:T} \mid y_{1:T}) \ \myd x_1 \myd w_{1:T}
  \end{aligned}
\end{equation}
which serves as a lower bound to the likelihood, given our current best estimate of the model parameters, $\theta_k$. With the joint log likelihood $\log p_\theta(y_{1:T},x_1,w_{1:T})$ given by (\ref{eq:jointDistLVN_prod}), to evaluate $Q(\theta,\theta_k)$ we must compute $p_{\theta_k}(x_1,w_{1:T} \mid y_{1:T})$, i.e., the joint smoothing distribution (JSD) of the initial state and disturbances. The E step then amounts to solving a \emph{disturbance smoothing}, rather than \emph{state smoothing}, problem, which reflects the use of disturbances, rather than states, as latent variables.

\par As demonstrated in Section \ref{sec:ZforDynamical}, $L_\theta(y_{1:T},x_1,w_{1:T})$ involves the simulated state sequence $\mathcal{X}_T(\theta,$\\$u_{1:T},x_1,w_{1:T})$. As a consequence, we shall show that the M step is equivalent to a nonconvex simulation error minimization problem (c.f. (\ref{eq:simErrorMinProb})). This is in contrast to the latent states formulation, in which $\max_\theta \ Q(\theta,\theta_k)$ reduces to linear least squares.

\subsection{Expectation step}\label{sec:EM_estep}
To compute $Q(\theta,\theta_k)$ it is convenient to use the following decomposition 
\begin{equation}\label{eq:logJointDistDecomp}
  \begin{aligned}
    Q(\theta,\theta_k)
      &= \underbrace{\CExpb{\theta_k}{\log p_{\theta}(x_1)}{y_{1:T}}}_{Q_1(\theta,\theta_k)} + \underbrace{\CExpb{\theta_k}{\log p_{\theta}(w_{1:T})}{y_{1:T}}}_{Q_2(\theta,\theta_k)} \\
      &\quad+ \underbrace{\CExpb{\theta_k}{\log p_{\theta}(y_{1:T} \mid x_1,w_{1:T})}{y_{1:T}}} _{Q_3(\theta,\theta_k)}
  \end{aligned}
\end{equation}
which was obtained by inserting (\ref{eq:jointDistLVN_prod}) into (\ref{eq:Q_lvn}).

\begin{rmk}\label{rem:Q_decomp}
Each term in (\ref{eq:logJointDistDecomp}) is a function of different parameters: $\mu$ and $\Sigma_1$ appear only in $Q_1(\theta,\theta_k)$; $\Sigma_w$ in $Q_2(\theta,\theta_k)$; and $\Sigma_v,A,B,G,C,D$ in $Q_3(\theta,\theta_k)$. To emphasize this we introduce the following decomposition of $\theta$
\begin{equation*}
\tha=\lbrace\mu,\Sigma_1\rbrace, \ \thb=\Sigma_w, \ \thc=\lbrace \Sigma_v,A,B,G,C,D\rbrace.
\end{equation*}
\end{rmk}
The following lemma details the computation of $Q(\theta,\theta_k)$. For clarity of expression, we introduce the following \emph{lifted} form of the dynamics in (\ref{eq:lgss}),
\begin{equation*}
Y = \bar{C}\bar{F}Z+(\bar{C}\bar{G}+\bar{D})U + V
\end{equation*}
where $Y = \text{vec}(y_{1:T})$, $U = \text{vec}(u_{1:T})$, $V = \text{vec}(v_{1:T})$, $Z = \text{vec}([x_1,w_{1:T-1}])$,  
\begin{equation*}
\bar{F}=\left[\begin{array}{cccccc}
I & 0 & 0 & 0 & \dots & 0 \\
A & G & 0 & 0 & \dots & 0 \\
A^2 & A & G & 0 & \dots & 0 \\
\vdots & & & & \ddots & \vdots \\
A^{T-1} & A^{T-2}G & A^{T-3}G & & \dots & G
\end{array}\right],
\end{equation*}
\begin{equation*}
\bar{G}=\left[\begin{array}{cccccc}
0 & 0 & 0 & 0 & \dots & 0 \\
B & 0 & 0 & 0 & \dots & 0 \\
AB & B & 0 & 0 & \dots & 0 \\
\vdots & & & & \ddots & \vdots \\
A^{T-2}B & A^{T-3}B & \dots & AB & B & 0
\end{array}\right],
\end{equation*}
$\bar{C}=I_T\otimes C$ and  $\bar{D}=I_T\otimes D$.

\begin{lem}
The auxiliary function $Q(\theta,\theta_k)$ defined in (\ref{eq:Q_lvn}) is given by
\begin{align*}
 Q(\theta,\theta_k) \propto& -\log\det\Sigma_1 - |\hat{x}_{1 \mid T}-\mu|_{\Sigma_1^{-1}}^2 - \textup{tr}\big( \Sigma_1^{-1}\widehat{\Sigma}_{1 \mid T} \big) \\
& -T\log\det\Sigma_w - \sum_{t=1}^{T}\textup{tr}\big(\Sigma_w^{-1}  \CExpb{\theta_k}{ w_t w_t'}{Y} \big) \\
& -T\log\det\Sigma_v - \textup{tr}(\Sigma_Y^{-1}(\bar{C}\bar{F}\Omega\bar{F}'\bar{C}' + \Delta\Delta'))
\end{align*}
where  
\begin{subequations}\label{eq:smoothedIC}
	\begin{align}
		\hat{x}_{1 \mid T} &= \CExpb{\theta_k}{x_1}{y_{1:T}}, \\
		\widehat{\Sigma}_{1 \mid  T} &= \CVarb{\theta_k}{x_1}{y_{1:T}}, 
\end{align}
\end{subequations}
\begin{subequations}\label{eq:dist_jsd}
	\begin{align}
		\hat{Z} & =\textup{E}_{\theta_k}\left[Z \mid y_{1:T}\right], \\
		\Omega & =\textup{Var}_{\theta_k}\left[Z \mid y_{1:T}\right],
	\end{align}
\end{subequations}
\begin{subequations}\label{eq:pyz_dist}
	\begin{align}
		\mu_Y &\triangleq \CExpb{\theta}{Y}{Z}=\bar{C}\bar{F}Z+(\bar{C}\bar{G}+\bar{D})U, \\
		\Sigma_Y &\triangleq \CVarb{\theta}{Y}{Z} = I_T\otimes\Sigma_v,
	\end{align}
\end{subequations}
\begin{equation}\label{eq:Delta}
	\Delta=\CExpb{\theta_k}{Y-\mu_Y}{y_{1:T}} = Y-\bar{C}\bar{F}\hat{Z}-(\bar{C}\bar{G}+\bar{D})U.
\end{equation}
\end{lem}

\begin{pf}
The first term in (\ref{eq:logJointDistDecomp}) is given by
\begin{align*}
  \CExpb{\theta_k}{\log p_{\theta}(x_1)}{y_{1:T}}
    &= \CExpb{\theta_k}{\log\mathcal{N}(x_1;\mu,\Sigma_1)}{y_{1:T}} \\
    &= \CExpb{\theta_k}{-\frac{n_x}{2}\log2\pi -\frac{1}{2}\log\det\Sigma_1 - |x_1-\mu|_{\Sigma_1^{-1}}^2}{y_{1:T}}.
\end{align*}
Ignoring constant terms and scaling factors yields
\begin{equation}\label{eq:Epx1_expand}
Q_1(\tha,\theta_k)\propto -\log\det\Sigma_1 - |\hat{x}_{1 \mid T}-\mu|_{\Sigma_1^{-1}}^2 - \textup{tr}\big( \Sigma_1^{-1}\widehat{\Sigma}_{1 \mid  T} \big) 
\end{equation}
where $\hat{x}_{1|T}$ and $\widehat{\Sigma}_{1|T}$ are given in (\ref{eq:smoothedIC}).

\par As the disturbances are i.i.d., $Q_2(\thb,\theta_k)$ is given by 
\begin{align*}
  \CExpb{\theta_k}{\log p_{\theta}(w_{1:T})}{y_{1:T}}
    &= \CExpb{\theta_k}{\log\prod_{t=1}^T\mathcal{N}(w_t;0,\Sigma_w)}{y_{1:T}} \\ 
    &= \sum_{t=1}^T \CExpb{\theta_k}{-\frac{n_w}{2}\log2\pi -\frac{1}{2}\log\det\Sigma_w - \frac{1}{2}|w_t|_{\Sigma_w^{-1}}^2 }{y_{1:T}}.
\end{align*}
Once more ignoring constants, this reduces to
\begin{align}\label{eq:LGSS_Q2}
Q_2(\thb,\theta_k)\propto -T\log\det\Sigma_w - \sum_{t=1}^{T}\textup{tr}\big(\Sigma_w^{-1}  \CExpb{\theta_k}{w_t w_t'}{y_{1:T}} \big).
\end{align}

\par Finally, we turn our attention to $Q_3(\thc,\theta_k)$. The p.d.f. $p_\theta(y_{1:T}\mid x_1,w_{1:T})$ is given by $p_\theta(Y \mid Z)=\mathcal{N}(Y;\mu_Y,\Sigma_Y)$, where $\mu_Y$ and $\Sigma_Y$ are given in (\ref{eq:pyz_dist}). $Q_3(\thc,\theta_k)$ may then be expressed as
\begin{align*}
  \CExpb{\theta_k}{\log\mathcal{N}(Y;\mu_Y,\Sigma_Y)}{y_{1:T}} 
    &= -\frac{Tn_y}{2} \log2\pi - \log\det\Sigma_Y - \CExpb{\theta_k}{|Y-\mu_Y|_{\Sigma_Y^{-1}}^2}{y_{1:T}}.
\end{align*}
Letting $\hat{Z}$ and $\Omega$, defined in (\ref{eq:dist_jsd}), denote the mean and covariance (respectively) of $p_{\theta_k}(x_1,$\\$w_{1:T-1}|y_{1:T})$, gives
\begin{equation}\label{eq:Q3_simple}
Q_3(\thc,\theta_k) \propto -T\log\det\Sigma_v - \text{tr}(\Sigma_Y^{-1}(\bar{C}\bar{F}\Omega\bar{F}'\bar{C}' + \Delta\Delta'))
\end{equation}
where $\Delta=\CExpb{\theta_k}{Y-\mu_Y}{y_{1:T}}$ is defined in (\ref{eq:Delta}). 
\end{pf}

Calculating the quantities in (\ref{eq:smoothedIC}) amounts to a state smoothing problem, the solution for which is given in closed form by, e.g., the RTS smoother \cite{rauch1965maximum} (see also, \cite[Section 4.4]{durbin2012time}). Similarly, for the LGSS models considered in this work, $\CExpb{\theta_k}{w_tw_t'}{Y_T}$, $\hat{Z}$ and $\Omega$ can be computed in closed form by standard disturbance smoothers; see, e.g., \cite[Section 4.5]{durbin2012time}.

\subsection{Maximization step}\label{sec:EM_mstep}
To perform the M step, i.e. maximize $Q(\theta,\theta_k)$, we will utilize the same decomposition as in (\ref{eq:logJointDistDecomp}), and optimize each of the conditional expectations separately; the validity of this approach is established by Remark \ref{rem:Q_decomp}.  
\par We begin with maximization of $Q_1(\tha,\theta_k)$:
\begin{lem}
The solution to $\tha_{k+1}=\arg\max_{\tha} \ Q_1(\tha,\theta_k)$
is given by $\tha_{k+1}=\lbrace \hat{x}_{1|T}, \widehat{\Sigma}_{1|T}\rbrace$.
\end{lem}
\begin{pf}
\par To maximize $Q_1(\tha,\theta_k)$, notice that (\ref{eq:Epx1_expand}) is concave w.r.t. $\mu$ and $\Sigma_1^{-1}$. Therefore, setting the gradient to zero gives the global maximizers $\mu=\hat{x}_{1|T}$ and $\Sigma_1=\widehat{\Sigma}_{1|T}$.
\end{pf}
Maximization of $Q_2(\thb,\theta_k)$ can be handled in a similar way: 
\begin{lem}
The solution to $\thb_{k+1}=\arg\max_{\thb} \ Q_2(\thb,\theta_k)$ is given by $\thb_{k+1}=\hat\Sigma_{w}$ where
\begin{equation}\label{eq:smoothNoiseCov}
\hat\Sigma_{w}=\frac{1}{T}\sum_{t=1}^{T} \CExpb{\theta_k}{w_t w_t'}{y_{1:T}}.
\end{equation}
\end{lem}
\begin{pf}
Substituting (\ref{eq:smoothNoiseCov}) into (\ref{eq:LGSS_Q2}) yields 
\begin{equation*}
Q_2(\thb,\theta_k)\propto-T\log\det\Sigma_w-T\textup{tr}\big(\Sigma_w^{-1}\hat\Sigma_{w} \big).
\end{equation*} 
This function is concave w.r.t. $\Sigma_w$ and so setting the gradient to zero gives the global maximizer $\Sigma_w=\hat\Sigma_{w}$. 
\end{pf}

\par Finally, we consider maximization of $Q_3(\thc,\theta_k)$. This is a challenging problem, due to its dependence on \emph{simulated} state sequences; c.f. Section \ref{sec:ZforDynamical}. Indeed, from (\ref{eq:Q3_simple}), it is clear that the quantities $\bar{F}$ and $\bar{G}$ render $Q_3(\thc,\theta_k)$ a nonconvex function of the model parameters.  

\par To maximize $Q_3(\thc,\theta_k)$ it is convenient to conceptualize (\ref{eq:Q3_simple}) as the summation of $T+1$ simultaneous simulation error minimization problems. 
\begin{lem}\label{lem:Q3}
Recalling the definition of simulation error in (\ref{eq:sim_error}), $Q_3(\thc,\theta_k)$ in (\ref{eq:logJointDistDecomp}) is equivalent to:
\begin{align}\label{eq:Q3_simerror}
-Q_3(\thc,\theta_k) = \ &\mathcal{E}(\theta,u_{1:T},y_{1:T},\hat{x}_{1|T},\hat{w}_{1:T}) + \sum_{j=1}^T \mathcal{E}(\theta,0,0,x_1^j,{w}^j_{1:T}) + T\log\det\Sigma_v
\end{align}
where $x_1^j,{w}^j_{1:T}$ are such that $\Omega = \sum_{j=1}^{T}\omega_j\omega_j'$ for $\omega_j=\textup{vec}([x_1^j,w_{1:T-1}^j])$.
\end{lem}

\begin{pf}
First consider the $\text{tr}(\Sigma_Y\Delta\Delta')$ term in (\ref{eq:Q3_simple}). From (\ref{eq:Delta}), $\Delta$ is clearly the difference between the measured output $y_{1:T}$ and the simulated output of the model with the expected value of the latent disturbances, i.e.      
\[\lbrace \hat{x}_{1|T}, \hat{w}_{1:T-1}\rbrace = \CExpb{\theta_k}{x_1,w_{1:T-1}}{y_{1:T}}=\hat{Z}.\] Therefore,
\begin{equation*}
\text{tr}(\Sigma_Y\Delta\Delta') = \sum_{t=1}^{T}|y_t-Cx_t-Du_t|_{\Sigma_v^{-1}}^2
\end{equation*}
where $x_{t+1}=Ax_t+Bu_t+G\hat{w}_t$ with $x_1=\hat{x}_{1|T}$.

\par Next, consider the $\text{tr}(\Sigma_Y^{-1}\bar{C}\bar{F}\Omega\bar{F}'\bar{C}')$ term. Decomposing $\Omega$ as the sum of $T$ rank one matrices, i.e. $\Omega = \sum_{j=1}^{T}\omega_j\omega_j'$, leads to
\begin{equation*}
\text{tr}(\Sigma_Y^{-1}\bar{C}\bar{F}\Omega\bar{F}'\bar{C}') = \sum_{j=1}^{T}|\bar{C}\bar{F}w_j|_{\Sigma_Y^{-1}}^2 = \sum_{j=1}^{T}\sum_{t=1}^{T}|Cx_t^j|_{\Sigma_v^{-1}}^2
\end{equation*}
where $x_{t+1}^j=Ax_t^j+Gw_t^j$. One can interpret this as the sum of $T$ simulation error problems with $y_{1:T}\equiv0$, $u_{1:T}\equiv0$, $\lbrace x_1,w_{1:T}\rbrace = \omega_j$. 
\end{pf}

To summarize, the computations involved in each iteration of the EM algorithm (formulated with latent disturbances) are straightforward, with the exception of maximization of $Q_3(\thc,\theta_k)$. From (\ref{eq:Q3_simerror}), this maximization is equivalent to $T+1$ simultaneous nonconvex simulation error minimization problems.

\section{Lagrangian relaxation of maximization step}\label{sec:lr_mstep}
In this section we describe how Lagrangian relaxation (c.f. Section \ref{sec:LRintro}) can be applied to the optimization of $Q_3(\thc,\theta_k)$ in (\ref{eq:Q3_simerror}). Specifically, we shall develop a bound for $Q_3(\thc,\theta_k)$ that can be efficiently optimized as a convex program. Furthermore, we shall show how this approach naturally enforces model stability at each iteration, by searching over a convex parametrization of all stable linear models.

\subsection{Lagrangian relaxation of simulation error}
Optimization of $Q_3(\thc,\theta_k)$ in (\ref{eq:Q3_simerror}) is difficult because it requires minimization of \emph{simulation error}, as defined in (\ref{eq:sim_error}). In this subsection, we detail the application of Lagrangian relaxation, introduced in Section \ref{sec:LRintro}, to minimization of simulation error, which can be formulated as
\begin{subequations}\label{eq:simErrorMinProb}
\begin{align}
J^* \triangleq \min_{\searg,x_{1:T}} \ & J(\searg,x_{1:T})\triangleq\sum_{t=1}^{T}\vert y_t-Cx_t-Du_t\vert_{\Sigma_v^{-1}}^2 \\
\textup{s.t.} \ & \mathcal{F}(\searg,u_{1:T},\initstate,x_{1:T},w_{1:T})=0. \label{eq:simErrorProb_dyn}
\end{align} 
\end{subequations}
Here $\initstate$ denotes the initial state, assumed to be known, and $\mathcal{F}(\searg,u_{1:T},\initstate,x_{1:T},w_{1:T})$ encodes the dynamic constraints on $x_{1:T}$, such that 
\[\mathcal{F}(\searg,u_{1:T},\initstate,\mathcal{X}_T(\theta,u_{1:T},\initstate,w_{1:T}),w_{1:T})=0.\]
As in (\ref{eq:LRgeneral}), the Lagrangian relaxation of (\ref{eq:simErrorMinProb}) takes the form 
\begin{align}\label{eq:Jsup}
& \lagj_{\mult}(\searg,u_{1:T},y_{1:T},\initstate,w_{1:T}) \triangleq \sup_{x_{1:T}} \ \lbrace J(\searg,x_{1:T})-\mult'\mathcal{F}(\searg,u_{1:T},\initstate,x_{1:T},w_{1:T})\rbrace,
\end{align}
and, for arbitrary multiplier $\mult$, represents a convex upper bound on the simulation error $\mathcal{E}(\searg,u_{1:T},$\\$y_{1:T},x_1,w_{1:T})$.

\subsection{Implicit dynamics}\label{sec:implicit_dynamics}
\par It remains to choose the Lagrange multiplier $\mult$ such that $\lagj_{\mult}(\searg)$ is a useful upper bound, i.e., such that $\lagj_{\mult}^*\approx J^*$. Unfortunately, the simultaneous search for $\mult$ and $\searg$ is not jointly convex, due to the coupling between $\mult$ and $\mathcal{F}$, and so $\mult$ must be specified in advance. However, we can alleviate this restriction by searching over an implicit representation of the dynamics in (\ref{eq:lgss_dynamics})
\begin{subequations}\label{eq:lgss_implicit}
\begin{align}
Ex_{t+1} &=Fx_t+Ku_t+Lw_t \label{eq:lgss_implicit_dyn}, \\
y_t &= Cx_t + Du_t + v_t,
\end{align}
\end{subequations}
where $E$ is invertible such that $A=E^{-1}F$, $B=E^{-1}K$ and $G=E^{-1}L$. With the implicit dynamics of (\ref{eq:lgss_implicit}), the dynamics constraint can be expressed 
\begin{equation}\label{eq:implicit_constraint}
\mathcal{F}(\eta,u_{1:T},\initstate,x_{1:T},w_{1:T}) = \bar{F}\text{vec}(x_{1:T})+\epsilon
\end{equation}
where $\bar{F}\in\mathbb{R}^{Tn_x\times Tn_x}$ and $\epsilon\in\mathbb{R}^{Tn_x}$ denote
\begin{equation*}
\left[\begin{array}{ccccc}
E & 0 & \dots\\
-F & E & 0 \\
0 & -F & E & 0 &  \\
\vdots & & & \ddots & \ddots
\end{array}\right] \ \textup{\&} \
\left[\begin{array}{c}
-E\initstate \\
K{u}_{1}+Lw_1 \\
\vdots \\
K{u}_{T-1}+Lw_{T-1}
\end{array}\right]
\end{equation*}
respectively. One may interpret the convex bound resulting from this implicit formulation as that of (\ref{eq:Jsup}), but with the multiplier $(I\otimes E')\mult$, thereby allowing a simultaneous (partial) search for multipliers and model parameters.

\begin{rmk}
We introduce $\eta=\lbrace E,F,K,L,C,D,\Sigma_v,P\rbrace$ to group the implicit model parameters, $\Sigma_v$ and $P\in\mathbb{S}_{++}^{n_x}$, into a single variable. Here $P$ represents a model stability certificate, the role of which is made precise in Lemma \ref{lem:stableModel}. Henceforth, $\lagj_\mult(\eta)$ denotes Lagrangian relaxation with the implicit dynamics constraint (\ref{eq:implicit_constraint}). For convenience, we define the mapping $\mathcal{M}:\eta\mapsto\gamma$ from an implicit to explicit parametrization: $\mathcal{M}(\eta)\triangleq \lbrace \Sigma_v, E\backslash F, E\backslash K, E\backslash L,C,D\rbrace$.
\end{rmk}

\subsection{Convex upper bound for $-Q_3(\thc,\theta_k)$ }
The representation of $Q_3(\thc,\theta_k)$ in (\ref{eq:Q3_simerror}) makes the application of Lagrangian relaxation straightforward. To obtain a convex upper bound for $-Q_3(\thc,\theta_k)$ we can simply replace each simulation error term $\mathcal{E}(\searg)$ with the appropriate corresponding convex bound $\lagj_\mult(\eta)$. 
\begin{lem}
Consider the following function
\begin{equation}\label{eq:bound_on_Q3}
\begin{aligned}
  \qbnd(\eta)
    &\triangleq \lagj_{\mult^0}(\eta,u_{1:T},y_{1:T},\hat{x}_{1|T},\hat{w}_{1:T}) + \sum_{j=1}^{T}\lagj_{\mult^j}(\eta,0,0,x_1^j,w_{1:T}^j) \\
    &\quad+ T\textup{tr}\Sigma_{v_k}^{-1}\Sigma_v + T\log\det\Sigma_{v_k} + Tn_y,
\end{aligned}
\end{equation}
where $x_1^j,w_{1:T}^j$ are defined in Lemma \ref{lem:Q3}.
$\qbnd(\eta)$ is a convex upper bound for $-Q_3(\thc,\theta_k)$, where $\thc=\mathcal{M}(\eta)$.
\end{lem}
\begin{pf}
As $\qbnd(\eta)$ is defined by a summation of convex functions, it is itself a convex function. Summation of the following inequalities 
\begin{align*}
\lagj_{\mult^0}(\eta,u_{1:T},y_{1:T},\hat{x}_{1|T},\hat{w}_{1:T}) &\geq \mathcal{E}(\searg,u_{1:T},y_{1:T},\hat{x}_{1|T},\hat{w}_{1:T}), \\
\lagj_{\mult^j}(\eta,0,0,x_1^j,w_{1:T}^j) &\geq \mathcal{E}(\searg,0,0,x_1^j,{w}^j_{1:T}), \quad j=1,\dots,T, \\
\text{tr}(\Sigma_{v_k}^{-1}\Sigma_v) + \log\det\Sigma_{v_k} + n_y &\geq \log\det\Sigma_v,
\end{align*}
gives $\qbnd(\eta)\geq -Q_3(\thc,\theta_k)$. Notice that $n_y + \log\det\Sigma_{v_k}+\text{tr}(\Sigma_{v_k}^{-1}\Sigma_v)$ is an affine upper bound on the concave term $\log\det\Sigma_v$, which is tight at our current best estimate of the covariance, $\Sigma_{v_k}$. 
\end{pf}

Notice that $\qbnd(\eta)$ is a function of an implicit representation of the dynamical model, denoted $\eta$, reflecting the fact that we formulate the Lagrangian relaxation using the implicit dynamics of (\ref{eq:lgss_implicit_dyn}).

\subsection{Lagrange multipliers}\label{sec:multipliers} 
To utilize the convex bound $\qbnd(\eta)$ we must supply suitable Lagrange multipliers, $\lbrace\mult^j\rbrace_{j=0}^T$. While convexity of the upper bound $\lagj_\mult$ defined in (\ref{eq:Jsup}) is guaranteed for any multiplier $\mult$ that is independent of $\eta$, in this work we consider multipliers of the form
$\mult=\text{vec}\left(\lbrace \mult_t\rbrace_{t=1}^T\right)$ for $\mult_t=2\left(Hx_t+h_t\right)$, i.e.
\begin{equation}\label{eq:multiplier_ext_form}
\mult = 2\left(\Lambda\text{vec}(x_{1:T}) + h\right),
\end{equation}
where $\Lambda=I_T\otimes H$ for $H\in\mathbb{R}^{n_x\times n_x}$ and $h\in\mathbb{R}^{Tn_x}$. Recall from Section \ref{sec:implicit_dynamics} that the use of the implicit model class (\ref{eq:lgss_implicit}) allows a convex (partial) search over model parameters and multipliers. Furthermore, this implicit representation permits the following definition of a convex parametrization of all stable LTI models.   

\begin{lem}\label{lem:stableModel}
Let $\Theta(H)$ denote the set of all models $\eta$ of the form (\ref{eq:lgss_implicit}) and $P\in\mathbb{S}_{++}^{n_x}$ that satisfy the LMI  
\begin{equation}\label{eq:stabilityCond}
M(\eta,H)=\left[\begin{array}{ccc}
H'E + E'H - P & F'H & C' \\
H'F & P & 0 \\
C & 0 & \Sigma_v
\end{array}\right] > 0 
\end{equation}
i.e. $\Theta(H)\triangleq\lbrace \eta: M(\eta,H)>0\rbrace.$
\par Then a model $\theta$ of the form (\ref{eq:lgss}) is stable iff there exists $E$ such that $\eta=\lbrace E,EA,EB,EG,C,D,P\rbrace\in\Theta(H)$ for some full-rank $H\in\real^{n_x\times n_x}$.
\end{lem}
\begin{pf}
This result is a straightforward extension of Lemma 4 and Corollary 5 in \cite[Section 3.2]{Manchester2012}.
\end{pf} 
\begin{rmk}\label{rem:E_invert}
The LMI $M(\eta,H)>0$ implies $H'E+E'H>0$ which ensures that $E$ is invertible, i.e., the implicit dynamics in (\ref{eq:lgss_implicit}) are well-posed. 
\end{rmk}

The model stability condition (\ref{eq:stabilityCond}) and multiplier (\ref{eq:multiplier_ext_form}) also guarantee finiteness of the supremum in (\ref{eq:Jsup}):
\begin{lem}\label{lem:finite_sup}
Given arbitrary $u_{1:T}$, $y_{1:T}$, $\initstate$, $w_{1:T}$, $h$ and full-rank $H$ the supremum in the definition of $\lagj_\mult(\eta)$ given by (\ref{eq:Jsup}) is finite, for $\mult$ given by (\ref{eq:multiplier_ext_form}) and $\eta\in\Theta(H)$.
\end{lem}
\begin{pf}
For ease of exposition, we define 
\begin{equation}\label{eq:Jmult}
J_\mult(\eta,x_{1:T}) \triangleq J(\eta,x_{1:T})-\mult'\mathcal{F}(\eta,u_{1:T},\initstate,x_{1:T},w_{1:T}),
\end{equation}
where $u_{1:T}$, $w_{1:T}$, $\initstate$ are dropped from the notation for brevity.
The bound in (\ref{eq:Jsup}) may then be equivalently expressed as $\lagj_\mult(\eta)=\sup_{x_{1:T}} J_\mult(\eta,x_{1:T})$. We can write 
\begin{equation}\label{eq:Jquad}
  J_\lambda(\eta,x_{1:T})
    =\sum_{t=1}^{T}x_t'C'\Sigma_v^{-1}Cx_t 
-2\sum_{t=1}^{T-1}(Hx_{t+1})'(Ex_{t+1}-Fx_t)
-2(Hx_1)'Ex_1 + \textup{aff}(x_{1:T})
\end{equation}
where $\text{aff}(x_{1:T})$ denotes additional terms that are affine in $x_{1:T}$.
We make use of the inequality 
\[ 2a'b\leq |a|^2_P + |b|^2_{P^{-1}} \quad \forall \ a,b,P>0 \]
(see, e.g., \cite[Section IV]{Tobenkin2010}) to obtain an upper bound for $J_\mult(\eta,x_{1:T})$. Specifically, by setting $a=x_{t+1}$ and $b = H'Fx_t$ we obtain the inequality
\begin{equation}\label{eq:keyIneq}
2x_{t+1}'H'Fx_t\leq |x_{t+1}|_P^2 + |H'Fx_{t}|_{P^{-1}}^2
\end{equation}
which holds for all $x_t,x_{t+1}$. Applying to (\ref{eq:keyIneq}) to (\ref{eq:Jquad}) yields the following upper bound:
\begin{equation}\label{eq:upBndJmult}
\begin{aligned}
  J_\mult(\eta,x_{1:T}) 
    &\leq x_1'\left(|H'F|^2_{P^{-1}}+C'\Sigma_v^{-1}C\right)x_1 \\
    &\quad+ \sum_{t=2}^{T-1}x_t'\left(|H'F|^2_{P^{-1}}+P-2H'E+|C|^2_{\Sigma_v^{-1}} \right)x_t \\
    &\quad+ x_T'\left(P-2H'E+|C|^2_{\Sigma_v^{-1}}\right)x_T + \text{aff}(x_1:T).
\end{aligned}
\end{equation}
The supremum w.r.t. $x_{1:T}$ of the upper bound on the LHS of (\ref{eq:upBndJmult}) is finite when the quadratic component is concave, i.e., $|H'F|^2_{P^{-1}}+P-H'E-E'H+|C|^2_{\Sigma_v^{-1}}<0$. By the Schur complement this condition is equivalent to the LMI in (\ref{eq:stabilityCond}). As finiteness of the bound implies finiteness of $J_\mult(\eta,x_{1:T})$ this completes the proof. 
\end{pf}

\par The key to the EM algorithm is (\ref{eq:dLgdQ}), i.e., increasing $Q(\theta,\theta_k)$ guarantees an improvement in $L_{\theta}(y_{1:T})$. Consequently, we must ensure that optimization of $\qbnd(\eta)$ does not decrease ${Q}_3(\thc,\theta_k)$. This property holds if there exist multipliers $\lbrace\mult^j\rbrace_{j=0}^T$ such that the bound $\qbnd(\eta)$ is `tight' to $-Q_3(\thc,\theta_k)$ at $\thc=\mathcal{M}(\eta)=\thc_k$, and may be understood as an application of the MM principle of Section \ref{sec:EM}. 

\par To obtain such a set of multipliers, we can minimize the bound $\qbnd(\eta_k)$ w.r.t. the multipliers $\lbrace\mult^j\rbrace_{j=1}^T$ for a fixed $\eta_k$. Here $\eta_k$ is such that $\thc_k=\mathcal{M}(\eta_k)$. We propose a two-stage approach: 

\begin{enumerate}[i.]
\item For each of the $T+1$ bounds $\lagj_{\mult^j}(\eta)$ that comprise $\qbnd(\eta)$, solve the convex optimization problem
\begin{equation}\label{eq:mult_H}
H_j = \arg\min_{\Phi_j} \ \lagj_{\phi_j}(\eta_k) \text{ s.t. } \eta_k\in\Theta(\Phi_j),
\end{equation}
where $\phi_j = I_T\otimes \Phi_j\text{vec}(x_{1:T})$.
\item Set $\mult^j = 2\left(I_T\otimes H_j\text{vec}(x_{1:T}) + h_j\right)$ such that $\lagj_{\mult^j}(\eta_k)=\mathcal{E}(\searg_k)$, where $h_j$ is computed as in Lemma \ref{lem:multiplier}.
\end{enumerate}

\begin{lem}\label{lem:multiplier} 
Given a model $\eta_k\in\Theta(H)$ of the form (\ref{eq:lgss_implicit}), and arbitrary $u_{1:T}$, $y_{1:T}$, $x_1$, $w_{1:T}$ and full-rank $H$, let $\mult$ denote a multiplier of the form (\ref{eq:multiplier_ext_form}) with $h$ defined as
\begin{equation}\label{eq:hMult}
h = (\bar{F}')^{-1}\left(\Psi X^*-\bar{C}'\bar{\Sigma}_Y^{-1}(Y-\bar{D}U)+\Lambda'\epsilon\right).
\end{equation}
Furthermore, let $\theta_k$ be such that $A=E\backslash F$, $B=E\backslash K$ and $G=E\backslash L$.
With this multiplier $\lagj_\mult(\eta_k)=\mathcal{E}(\theta_k)$, i.e. the convex bound $\lagj_\mult(\eta)$ is tight to the simulation error $\mathcal{E}(\searg)$ at $\eta=\eta_k$. The notation is as follows: $\bar{F}$ and $\epsilon$ are defined in (\ref{eq:implicit_constraint}); $\bar{C}$, $\bar{D}$, $\Sigma_Y$, $U$, $Y$ are defined in Section \ref{sec:EM_estep}; $X^*=\textup{vec}(\mathcal{X}_T(\eta_k,u_{1:T},x_1,w_{1:T}))$ and 
\[\Psi= \bar{C}'\bar{\Sigma}_Y^{-1}\bar{C}-\Lambda'\bar{F}-\bar{F}'\Lambda.\]
\end{lem}

\begin{pf}
As $\eta_k\in\Theta$, by Lemma~\ref{lem:finite_sup}, we know that
$J_\mult(\eta,x_{1:T})$, defined in (\ref{eq:Jmult}),
is a concave quadratic function in $x_{1:T}$. By the first order optimality condition, it can be shown that the state sequence $x_{1:T}^*$ that maximizes this function must satisfy
\begin{equation*}
\Psi\text{vec}(x_{1:T}^*)=\bar{C}'(Y-\bar{D}U)+\bar{F}'h - \Lambda'\epsilon.
\end{equation*}
Setting $x_{1:T}^*=X^*$ and solving for $h$ yields the expression in (\ref{eq:hMult}). Note that invertibility of $E$ ensures that $\bar{F}^{-1}$ is well-defined; c.f. Remark \ref{rem:E_invert}. As $x_{1:T}^*=X^*$ we have
\begin{align*}
\lagj_\mult(\eta_k)&=J(\eta_k,X^*)-\mult\mathcal{F}(\eta_k,u_{1:T},X^*,w_{1:T}) =J(\eta_k,X^*)=\mathcal{E}(\eta_k). 
\end{align*} 

\end{pf}
In summary, to update $\thc$ at the $k\myth$ iteration of the EM algorithm, we solve the convex optimization problem
\begin{equation}\label{eq:Q3_bound_opt}
\eta_{k+1} = \arg\min_{\eta} \ \qbnd(\eta) \text{ s.t. } \eta\in\bigcap_{j=0}^T\Theta(H_j), 
\end{equation}
where $\mult^j$ is given by (\ref{eq:multiplier_ext_form}) with $H_j$ from (\ref{eq:mult_H}) and $h_j$ from (\ref{eq:hMult}), for $j=0,\dots,T$, and then set $\thc_{k+1}=\mathcal{M}(\eta_{k+1})$.
For a complete summary of EM with latent disturbances, refer to Algorithm \ref{alg:EMLVN}. 

\begin{rmk}
A common heuristic for terminating the EM algorithm is to cease iterations once the change in likelihood falls below a certain tolerance $\delta$, i.e.  
\begin{equation}\label{eq:term_heuristic}
L_{\theta_{k+1}}(y_{1:T})-L_{\theta_k}(y_{1:T})<\delta.
\end{equation}
Alternatively, one can simply run the algorithm for a finite number of iterations, chosen so as to attain a model of quality sufficient for its intended application; this is the approach taken, e.g., in \cite{gibson2005robust,wills2010estimating}.
\end{rmk}

\begin{algorithm}
\caption{EM with latent disturbances}\label{alg:EMLVN}
\begin{enumerate}
\item Set $k=0$ and initialize $\theta_k$ such that $L_{\theta_k}(y_{1:T})$ is finite.
\item \textbf{Expectation (E) Step:}
\begin{enumerate}[(2.1)]
\item Compute $\hat{x}_{1 \mid  T}$ and $\widehat{\Sigma}_{1 \mid  T}$ as in (\ref{eq:smoothedIC}).
\item Compute $\widehat{\Sigma}_{w}$ as in (\ref{eq:smoothNoiseCov}).
\item Compute $\hat{Z}$ and $\Omega$ as in (\ref{eq:dist_jsd}).
\end{enumerate}

\item \textbf{Maximization (M) Step:}
\begin{enumerate}[(3.1)]
\item Set $\tha_{k+1}=\lbrace \hat{x}_{1 \mid  T},\widehat{\Sigma}_{1 \mid T}\rbrace$ and $\thb_{k+1}=\widehat{\Sigma}_{w}$.
\item Assemble $\lbrace\mult^i\rbrace_{j=0}^T$ of the form (\ref{eq:multiplier_ext_form}) by computing $\lbrace H_j\rbrace_{j=0}^T$ with (\ref{eq:mult_H}) and $\lbrace h_j\rbrace_{j=0}^T$ with (\ref{eq:hMult}).
\item Compute $\eta_{k+1}$ by solving (\ref{eq:Q3_bound_opt}) and set $\thc_{k+1}=\mathcal{M}(\eta_{k+1})$.
\item Set $\theta_{k+1}=\lbrace\tha_{k+1},\thb_{k+1},\thc_{k+1}\rbrace$.
\end{enumerate}
\item Terminate if (\ref{eq:term_heuristic}), otherwise $k\gets k+1$ and return to step 2.
\end{enumerate}
\end{algorithm}

\section{Theoretical properties of identification via EM}

\subsection{Singular state space models}
In applications, it may arise that the dimension of the disturbance is less than that of the state variable, i.e. $n_w<n_x$. For example, consider a simple mass-spring-damper system governed by $m\ddot{s}+c\dot{s}+ks=u+w$ for displacement $s$. When discretized, these dynamics can be represented by the second order state space model
\begin{equation*}
x_{t+1}
= 
\left[\begin{array}{cc}
1 & \Delta_t \\
\frac{-k\Delta_t}{m} & 1 -\frac{c\Delta_t}{m}
\end{array}\right]
x_t
+
\left[\begin{array}{c}
0 \\
\Delta_t
\end{array}\right]u
+
\left[\begin{array}{c}
0 \\
\Delta_t
\end{array}\right]w
\end{equation*}
with state variable $x_t=\left[s(t) \ \dot{s}(t)\right]'$.

\par In such cases, the process noise covariance $G\Sigma_vG'$ is singular, and standard EM algorithms based on latent states are no longer applicable. To see why, observe that the transition density of such a model is given by
\begin{equation*}
p_\theta(x_{t+1} \mid x_t)=\mathcal{N}(x_{t+1};Ax_t+Bu_t,G\Sigma_wG').
\end{equation*}
As $G\Sigma_wG'$ is rank deficient, the transition $p_\theta(x_{t+1} \mid x_t)$ does not admit a density \cite{MullerStewart2006}, and so we cannot evaluate, much less optimize, the joint log likelihood $\log p_\theta(y_{1:T},x_{1:T})$ given in (\ref{eq:jointDistLVS}). Modifications to the standard latent states EM algorithm have been proposed to circumvent this difficulty e.g. the work of \cite{solo2003algorithm} introduces a perturbation model with full-rank process noise covariance. However, by choosing latent disturbances we can elegantly handle identification of both singular and full-rank state space models, with the same algorithm. 

\par In particular, when formulating the EM algorithm over latent disturbances we work with the joint likelihood function $p_\theta(y_{1:T},x_1,w_{1:T})$, given in (\ref{eq:jointDistLVN_prod}). Comparing (\ref{eq:jointDistLVN_prod}) to (\ref{eq:jointDistLVS}), we observe that the problematic transition density is replaced by the joint distribution of disturbances
\begin{equation*}
p_\theta(w_{1:T})=\prod_{t=1}^{T}\mathcal{N}(w_t;0,\Sigma_w).
\end{equation*}
This distribution is independent of $n_x$, and so $p_\theta(y_{1:T},x_1,w_{1:T})$ and, therefore, $Q(\theta,\theta_k)$ remains well-defined, even in the singular case, $n_w<n_x$.

\subsection{Absence of disturbances or output noise}\label{sec:limit_cases} 
In this section, we study the auxiliary function $Q\ththk$ in the limit cases of $\Sigma_w=0$ and $\Sigma_v=0$, for different choices of latent variables. These results will offer insight into the behavior of the EM algorithm as a function of disturbance magnitude, as explored in the numerical experiments of Section \ref{sec:1dbounds}. For convenience, we denote the bounds based on latent states and disturbances by $\qls\ththk$ and $\qld\ththk$, respectively.

\begin{prop}\label{prop:Qlvs_wSmall}
Consider a model of the form (\ref{eq:lgss}), and let $\theta$ be such that $\Sigma_w=0$, i.e. disturbances are omitted from the model. The auxiliary function built on \textup{latent states}, $\qls\ththk$, is undefined when $A\neq A_k$ or $B\neq B_k$. 
\end{prop}
\begin{pf}
When $\Sigma_w=0$, given any $x_1\in\real^{n_x}$ the p.d.f. $p_{\theta_k}(x_{1:T} \mid y_{1:T})$ is nonzero on the set $\mathcal{S}(\theta_k)=\lbrace x_{1:T} : x_{1:T}=\mathcal{X}(\theta_k,u_{1:T},x_1,0)  \ \forall \ x_1\in\real^{n_x}\rbrace$. The auxiliary function may be expressed as
\begin{equation*}
\qls\ththk = \int_{\mathcal{S}(\theta_k)}\log p_\theta(x_{1:T},y_{1:T})p_{\theta_k}(x_{1:T} \mid y_{1:T}) \ \myd x_{1:T}.
\end{equation*}
As $\Sigma_w=0$, $p_\theta(x_{2:T} \mid x_1)$ is deterministic, evaluating to unity when $x_{1:T}=\mathcal{X}(\theta,u_{1:T},x_1,0)$, and zero otherwise. When $A\neq A_k$ or $B\neq B_k$, $\log p_\theta(x_{2:T} \mid x_1)=0$ for all $x_{1:T}\in\mathcal{S}(\theta_k)$, and so $\log p_\theta(x_{1:T},y_{1:T})$ is undefined. As a consequence, $\qls\ththk$ is undefined.

\par When $A=A_k$ and $B=B_k$, $p_\theta(x_{2:T} \mid x_1)=1$ for all $x\in\mathcal{S}(\theta_k)$ and so $\qls\ththk$ can be evaluated as usual.  
\end{pf}
 
\begin{prop}\label{prop:Qlvn_wSmall}
Consider a model of the form (\ref{eq:lgss}), and let $\theta$ be such that $\Sigma_w=0$, i.e. disturbances are omitted from the model. Furthermore, suppose $\Sigma_1=0$; i.e. the initial conditions $x_1=\mu$ are modeled without uncertainty.  Then  $L_\theta(y_{1:T},x_1)=\qld\ththk$ for all $\theta,\theta_k$; i.e., the auxiliary function built on \textup{latent disturbances}, $\qld\ththk$, reduces to the log likelihood. 
\end{prop}
\begin{pf}
As $\Sigma_w=0, \ \Sigma_1=0$ the p.d.f. $p_{\theta_k}(x_1,w_{1:T} \mid y_{1:T})$ is trivially deterministic, evaluating to unity when $x_1=\mu$ and $w_{1:T}\equiv0$, and evaluating to zero otherwise. Therefore
\begin{align*}
\qld\ththk = \log p_\theta(y_{1:T},\mu,0) = \log p_\theta(y_{1:T} \mid \mu).
\end{align*}
The log likelihood can be decomposed as
\begin{align*}
L_\theta(y_{1:T})&=\log \int p_\theta(y_{1:T},x_1) \myd x_1 \\ &= \log \int p_\theta(y_{1:T} \mid x_1)p_\theta(x_1) \myd x_1 = \log p_\theta(y_{1:T} \mid \mu),
\end{align*}
where the final equality follows from the fact that $p_\theta(x_1)$ is a $\delta$-function, at $x_1=\mu$. 
\end{pf}

\begin{prop}\label{prop:Qlvn_zSmall}
Consider a \textup{first order} model of the form (\ref{eq:lgss}), and let $\theta$ be such that $\Sigma_v=0$, i.e. output noise is omitted from the model. The auxiliary function built on \textup{latent disturbances}, $\qld\ththk$, is undefined for $\theta\neq\theta_k$, i.e. $Q(\theta,\theta_k)$ collapses to a single point at $\theta=\theta_k$.
\end{prop}
\begin{pf}
For a given $\theta$, let $x_{1:T}^\theta$ denote the unique state sequence that is `consistent' with the data, i.e. $x_{1:T}^\theta\triangleq\lbrace x_{1:T}: y_t=Cx_t+Du_t, t = 1,\dots,T\rbrace$. There is also a corresponding unique disturbance sequence, denoted $w_{1:T}^\theta=\lbrace w_{1:T}: x_{1:T}^\theta=\mathcal{X}(\theta,u_{1:T},x_1^\theta,w_{1:T})\rbrace$. 
\par As $\Sigma_v=0$, the p.d.f. $p_{\theta_k}(x_1,w_{1:T} \mid y_{1:T})$ is a $\delta$-function at $x_1=x^{\theta_k}_1$ and $w_{1:T}=w^{\theta_k}_{1:T}$. The auxiliary function is then given by
\begin{equation*}
\qld\ththk = \log p_\theta(y_{1:T},x^{\theta_k}_1,w^{\theta_k}_{1:T}).
\end{equation*} 
We can decompose $p_\theta(y_{1:T},x^{\theta_k}_1,w^{\theta_k}_{1:T})$ as in (\ref{eq:jointDistLVN_prod}). As $\Sigma_v=0$, the p.d.f. $p_\theta(y_{1:T} \mid x_1,w_{1:T})$ is also a $\delta$-function at $x_1=x^{\theta}_1$ and $w_{1:T}=w^{\theta}_{1:T}$. If $C\neq C_k$ or $D\neq D_k$ then $x_1^\theta\neq x_1^{\theta_k}$. Furthermore, if $A\neq A_k$, $B\neq B_k$ or $G\neq G_k$, then $\mathcal{X}(\theta,u_{1:T},x_1^\theta,w_{1:T}^\theta)\neq\mathcal{X}(\theta_k,u_{1:T},x_1^{\theta_k},w_{1:T}^{\theta_k})$. In both cases $p_\theta(y_{1:T} \mid x_1^{\theta_k},w^{\theta_k}_{1:T})=0$ and so $\qld\ththk$ is undefined.

\par When $\theta=\theta_k$, $p_\theta(y_{1:T} \mid x_1^{\theta_k},w^{\theta_k}_{1:T})=1$ and $\qld\ththk$ can be evaluated as usual. 
\end{pf}

\begin{prop}\label{prop:Qlvs_zSmall}
Consider a \textup{first order} model of the form (\ref{eq:lgss}), and let $\theta$ be such that $\Sigma_v=0$, i.e. output noise is omitted from the model. Let $\qls\ththk$ denote the auxiliary function built on \textup{latent states}, then:
\begin{enumerate}[i.]
\item $\qls\ththk$ is undefined for all $\theta$ such that $C\neq C_k$ or $D\neq D_k$.
\item $\qls\ththk=\loglik$ for all $\theta$ such that $C=C_k$ and $D=D_k$.  
\end{enumerate}
\end{prop}
\begin{pf}
For a given $\theta$, let $x_{1:T}^\theta$ denote the unique state sequence that is `consistent' with the data, i.e. $x_{1:T}^\theta\triangleq\lbrace x_{1:T}\colon y_t=Cx_t+Du_t, t = 1,\dots,T\rbrace$. As $\Sigma_v=0$, given $y_{1:T}$ both $p_{\theta_k}(x_{1:T} \mid y_{1:T})$ and $p_\theta(y_{1:T} \mid x_{1:T})$ are $\delta$-functions at $x_{1:T}=x^{\theta}_{1:T}$. The auxiliary function is then given by
\begin{equation*}
\qls\ththk = \log p_\theta(y_{1:T},x^{\theta_k}_{1:T}).
\end{equation*}
Let us now consider the two cases:
\begin{enumerate}[i.]
\item When $C\neq C_k$ or $D\neq D_k$, $x^{\theta}_{1:T}\neq x^{\theta_k}_{1:T}$ and so $p_\theta(y_{1:T}\mid x^{\theta_k}_{1:T})=0$. Therefore, $\qls\ththk$ is undefined.
\item When $C=C_k$ and $D=D_k$, $x^{\theta}_{1:T}=x^{\theta_k}_{1:T}$ and so 
\begin{equation*}
\qls\ththk = \log p_\theta(y_{1:T} \mid x^\theta_{1:T})p_\theta(x^\theta_{1:T}) = \log p_\theta(x^\theta_{1:T}).
\end{equation*}
The likelihood can be expressed as
\begin{align*}
  L_\theta(y_{1:T})
    = \log \int p_\theta(y_{1:T} \mid x_{1:T})p_\theta(x_{1:T}) \myd x_{1:T}
    = \log p_\theta(x^\theta_{1:T}),
\end{align*}
where the second inequality comes from the fact that $p_\theta(y_{1:T} \mid x_{1:T})$ is a $\delta$-function. Therefore, $L_\theta(y_{1:T})=\qls\ththk$. 
\end{enumerate} 
\end{pf}

\section{Numerical experiments}

\subsection{Influence of disturbance magnitude on bound fidelity}\label{sec:1dbounds}
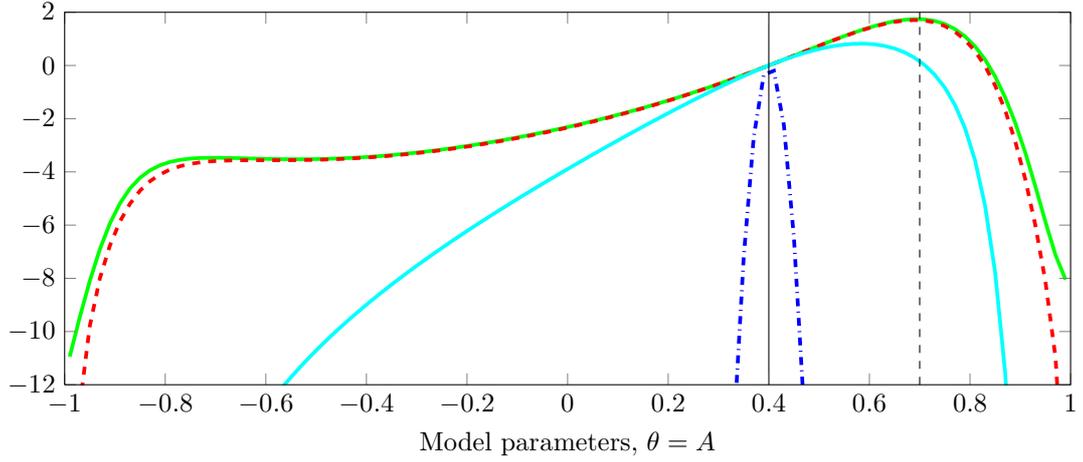
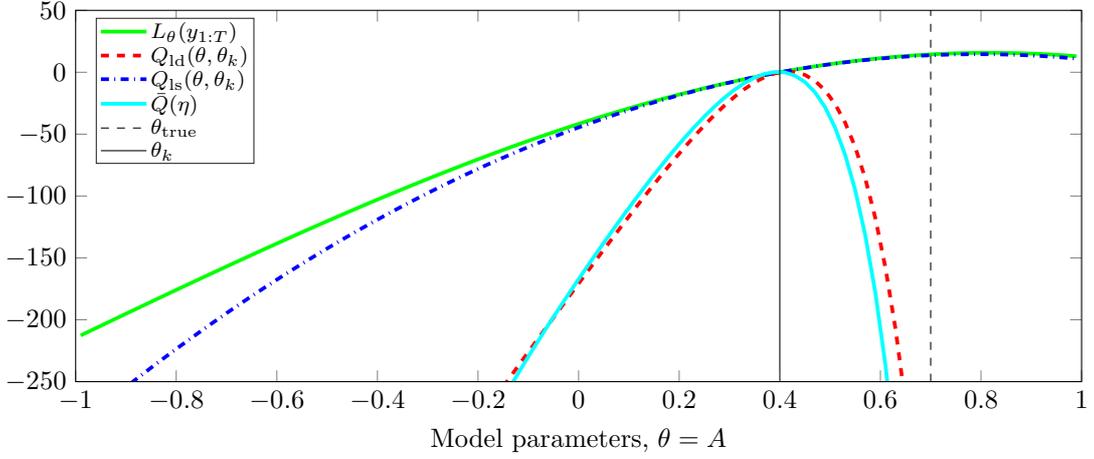
\begin{figure}
\centering
\subfloat[`Small' disturbances: $\Sigma_v=1\times10^{-3}$ and $\Sigma_v=1\times10^{-2}$.]
    	 {\setlength{\figureheight}{0.35\textwidth}\setlength{\figurewidth}{0.9\textwidth}
%
%
\pgfplotsset{
  axis on top,
  y tick label style={rotate=0},
  legend style={inner xsep=1pt,inner ysep=0.5pt,nodes={inner sep=1pt,text depth=0.1em},font=\footnotesize},
  minor y tick num=0,
  minor x tick num=0,
 }
\centering

\newcommand\minus{%
  -
}
\definecolor{mycolor1}{rgb}{0.00000,1.00000,1.00000}%
\begin{tikzpicture}

\begin{axis}[%
width=\figurewidth,
height=0.96\figureheight,
at={(0\figurewidth,0\figureheight)},
scale only axis,
xmin=-1,
xmax=1,
xtick={-1,-.8,-.6,-.4,-.2,0,.2,.4,.6,.8,1},
xticklabels={$\minus 1$,$\minus 0.8$,$\minus 0.6$,$\minus 0.4$,$\minus 0.2$,$0$,$0.2$,$0.4$,$0.6$,$0.8$,$1$},
xlabel={Model parameters, $\theta = A$},
ymin=-12,
ymax=2,
ytick={-12,-10,-8,-6,-4,-2,0,2},
yticklabels={$\minus 12$,$\minus 10$,$\minus 8$,$\minus 6$,$\minus 4$,$\minus 2$,$0$,$\phantom{-00}2$},
axis background/.style={fill=white}
]
\addplot [color=green,solid,line width=1.4pt,forget plot]
  table[row sep=crcr]{%
-0.99	-10.9444455970192\\
-0.97	-9.46552856626862\\
-0.95	-8.09903003481354\\
-0.93	-6.90986038620218\\
-0.91	-5.93732190926357\\
-0.89	-5.18270306276443\\
-0.87	-4.62149876657168\\
-0.85	-4.21837365009689\\
-0.83	-3.93744782351212\\
-0.81	-3.74743080918543\\
-0.79	-3.62322383066108\\
-0.77	-3.54567063496767\\
-0.75	-3.50058331477622\\
-0.73	-3.47763911556843\\
-0.71	-3.46940161544858\\
-0.69	-3.47053967638974\\
-0.67	-3.47723813564291\\
-0.65	-3.4867669328256\\
-0.63	-3.49717155086295\\
-0.61	-3.50705260257943\\
-0.59	-3.51540944518711\\
-0.57	-3.52152922713415\\
-0.55	-3.52490799329932\\
-0.53	-3.52519437621727\\
-0.51	-3.52214921170005\\
-0.49	-3.51561640319291\\
-0.47	-3.50550174996804\\
-0.45	-3.49175742593416\\
-0.43	-3.47437047592526\\
-0.41	-3.45335417430154\\
-0.39	-3.42874142829132\\
-0.37	-3.40057964807117\\
-0.35	-3.36892667615394\\
-0.33	-3.33384749024334\\
-0.31	-3.29541148024948\\
-0.29	-3.25369016144145\\
-0.27	-3.20875522874441\\
-0.25	-3.16067688703323\\
-0.23	-3.10952241265603\\
-0.21	-3.05535491508719\\
-0.19	-2.998232276617\\
-0.17	-2.9382062538618\\
-0.15	-2.87532172875643\\
-0.13	-2.80961609939119\\
-0.11	-2.74111880316514\\
-0.0900000000000001	-2.66985096664082\\
-0.0700000000000001	-2.59582517846235\\
-0.0499999999999999	-2.51904538388133\\
-0.03	-2.43950690190282\\
-0.01	-2.35719656883337\\
0.01	-2.27209301508142\\
0.03	-2.18416708540513\\
0.05	-2.09338241641326\\
0.0700000000000001	-1.99969618898883\\
0.0900000000000001	-1.90306007745045\\
0.11	-1.80342142172827\\
0.13	-1.70072465368921\\
0.15	-1.59491301411235\\
0.17	-1.48593060283268\\
0.19	-1.37372481144156\\
0.21	-1.25824919589095\\
0.23	-1.13946685571494\\
0.25	-1.01735439773239\\
0.27	-0.891906575534094\\
0.29	-0.76314171241313\\
0.31	-0.631108035498599\\
0.33	-0.49589107378398\\
0.35	-0.357622303933312\\
0.37	-0.216489267095568\\
0.39	-0.0727474299558111\\
0.41	0.0732658728465623\\
0.43	0.221114995796839\\
0.45	0.370246514563718\\
0.47	0.519966073293929\\
0.49	0.669410714771914\\
0.51	0.817515642785168\\
0.53	0.962974094115097\\
0.55	1.10418866640111\\
0.57	1.23921205823906\\
0.59	1.36567474279273\\
0.61	1.48069665413507\\
0.63	1.58077959907455\\
0.65	1.66167697549123\\
0.67	1.71823776803195\\
0.69	1.744223195676\\
0.71	1.73209761119502\\
0.73	1.67280156149717\\
0.75	1.55552615525228\\
0.77	1.36752649986194\\
0.79	1.09404073646742\\
0.81	0.718422410916119\\
0.83	0.222648320105563\\
0.85	-0.411567956729954\\
0.87	-1.20073111205662\\
0.89	-2.156057680953\\
0.91	-3.27735913452442\\
0.93	-4.54128706366112\\
0.95	-5.87667755274423\\
0.97	-7.12221114035326\\
0.99	-8.04933556557656\\
};
\addplot [color=red,dashed,line width=1.4pt,forget plot]
  table[row sep=crcr]{%
-0.97	-12.8064158899379\\
-0.95	-9.79248316928658\\
-0.93	-8.00772005182799\\
-0.91	-6.76868683557311\\
-0.89	-5.86310612563341\\
-0.87	-5.19461089684464\\
-0.85	-4.70365472643837\\
-0.83	-4.34683058809603\\
-0.81	-4.0907392276888\\
-0.79	-3.90956012765511\\
-0.77	-3.78352892215634\\
-0.75	-3.69768737286213\\
-0.73	-3.6408135079179\\
-0.71	-3.60453548529279\\
-0.69	-3.58262606789734\\
-0.67	-3.57045798285967\\
-0.65	-3.56459227231451\\
-0.63	-3.56247125309793\\
-0.61	-3.56219119749974\\
-0.59	-3.56233460418849\\
-0.57	-3.56184652110886\\
-0.55	-3.55994325132377\\
-0.53	-3.55604481283012\\
-0.51	-3.54972482161435\\
-0.49	-3.54067316815994\\
-0.47	-3.52866810302581\\
-0.45	-3.51355525504539\\
-0.43	-3.4952317672755\\
-0.41	-3.47363421865484\\
-0.39	-3.44872935256177\\
-0.37	-3.42050689251667\\
-0.35	-3.38897391564157\\
-0.33	-3.35415039455606\\
-0.31	-3.31606562149054\\
-0.29	-3.27475530426307\\
-0.27	-3.23025917954246\\
-0.25	-3.18261902978597\\
-0.23	-3.13187702031178\\
-0.21	-3.07807429506121\\
-0.19	-3.02124978589578\\
-0.17	-2.96143920237733\\
-0.15	-2.89867417809316\\
-0.13	-2.83298155659676\\
-0.11	-2.76438280558034\\
-0.0900000000000001	-2.69289355244078\\
-0.0700000000000001	-2.61852323827807\\
-0.0499999999999999	-2.54127489079752\\
-0.03	-2.46114501973955\\
-0.01	-2.37812364143687\\
0.01	-2.29219444198085\\
0.03	-2.20333509132553\\
0.05	-2.11151772352741\\
0.0700000000000001	-2.01670960127316\\
0.0900000000000001	-1.91887398595796\\
0.11	-1.81797123794271\\
0.13	-1.71396017535793\\
0.15	-1.60679972409659\\
0.17	-1.49645089664753\\
0.19	-1.38287914342536\\
0.21	-1.26605712757461\\
0.23	-1.14596798327992\\
0.25	-1.02260912891907\\
0.27	-0.895996720613056\\
0.29	-0.766170849695044\\
0.31	-0.63320161041068\\
0.33	-0.497196193144191\\
0.35	-0.358307195399703\\
0.37	-0.216742389912383\\
0.39	-0.0727762495254467\\
0.41	0.0732363944370604\\
0.43	0.220844084904456\\
0.45	0.369479261273312\\
0.47	0.51843480819943\\
0.49	0.666835662570946\\
0.51	0.81360424956867\\
0.53	0.957418190170728\\
0.55	1.0966583123533\\
0.57	1.22934449334946\\
0.59	1.35305625109956\\
0.61	1.4648342897797\\
0.63	1.56105840867298\\
0.65	1.63729636914613\\
0.67	1.68811762208506\\
0.69	1.70686550628159\\
0.71	1.6853821495537\\
0.73	1.61368273223468\\
0.75	1.47958148273659\\
0.77	1.2682830521461\\
0.79	0.961972978031525\\
0.81	0.539473575656103\\
0.83	-0.0239209100954838\\
0.85	-0.756260249435115\\
0.87	-1.68821250509407\\
0.89	-2.85178113154728\\
0.91	-4.27947338832786\\
0.93	-6.00942974036033\\
0.95	-8.1231514812799\\
0.97	-10.9766068749118\\
0.978021763619425	-13.4\\
};
\addplot [color=blue,dashdotted,line width=1.4pt,forget plot]
  table[row sep=crcr]{%
0.331608792313016	-13.4\\
0.35	-7.21699464507481\\
0.37	-2.68573689056655\\
0.39	-0.347092331361864\\
0.41	-0.201060967463206\\
0.43	-2.24764279887006\\
0.45	-6.48683782558248\\
0.47	-12.9186460475985\\
};
\addplot [color=mycolor1,solid,line width=1.4pt,forget plot]
  table[row sep=crcr]{%
-0.57	-12.1563380718012\\
-0.55	-11.716282209662\\
-0.53	-11.2990522198349\\
-0.51	-10.9019741187898\\
-0.49	-10.5227768714499\\
-0.47	-10.1595171641516\\
-0.45	-9.81052015612804\\
-0.43	-9.47433459013586\\
-0.41	-9.14969623707508\\
-0.39	-8.83549465102512\\
-0.37	-8.53075544374426\\
-0.35	-8.23461391660189\\
-0.33	-7.94630262411901\\
-0.31	-7.66513834717811\\
-0.29	-7.39050847710246\\
-0.27	-7.12186403348498\\
-0.25	-6.85870988073767\\
-0.23	-6.60060001231628\\
-0.21	-6.34713164863329\\
-0.19	-6.09793945807958\\
-0.17	-5.8526938272912\\
-0.15	-5.61109602444247\\
-0.13	-5.37287490225792\\
-0.11	-5.13778711497122\\
-0.0900000000000001	-4.90561416334149\\
-0.0700000000000001	-4.67616031157166\\
-0.0499999999999999	-4.44925291690014\\
-0.03	-4.22474087748792\\
-0.01	-4.00249491306988\\
0.01	-3.78240709572223\\
0.03	-3.56439139177131\\
0.05	-3.34838507221371\\
0.0700000000000001	-3.13434789431087\\
0.0900000000000001	-2.9222653927945\\
0.11	-2.71214968470834\\
0.13	-2.50404146014383\\
0.15	-2.29801322845086\\
0.17	-2.09417214982198\\
0.19	-1.8926626375891\\
0.21	-1.69367222901369\\
0.23	-1.49743438649421\\
0.25	-1.30423560021649\\
0.27	-1.11442039618144\\
0.29	-0.928397176376315\\
0.31	-0.746647804735002\\
0.33	-0.569733652703338\\
0.35	-0.398305547656967\\
0.37	-0.233112519164109\\
0.39	-0.075013241862429\\
0.41	0.0750132418624254\\
0.43	0.215855318071522\\
0.45	0.346253927111622\\
0.47	0.464786918514488\\
0.49	0.569847939936963\\
0.51	0.65962097655293\\
0.53	0.732046933538829\\
0.55	0.784770955793849\\
0.57	0.815075163202845\\
0.59	0.81977528478545\\
0.61	0.795068419710063\\
0.63	0.736305462190291\\
0.65	0.637652997309086\\
0.67	0.491580921571131\\
0.69	0.288067662365631\\
0.71	0.0133593566895414\\
0.73	-0.352003124984385\\
0.75	-0.836410513872814\\
0.77	-1.48254473337883\\
0.79	-2.35644412089127\\
0.81	-3.56305340428332\\
0.83	-5.27337136275023\\
0.85	-7.77334799794603\\
0.87	-11.5573487698251\\
0.876176408297381	-13.4\\
};
\addplot [color=black,dashed,forget plot]
  table[row sep=crcr]{%
0.7	-12\\
0.7	2\\
};
\addplot [color=black,solid,forget plot]
  table[row sep=crcr]{%
0.4	-12\\
0.4	2\\
};
\end{axis}
\end{tikzpicture}%
} \\
\subfloat[`Large' disturbances: $\Sigma_w=10$ and $\Sigma_v=1\times10^{-2}$.]
         {\setlength{\figureheight}{0.35\textwidth}\setlength{\figurewidth}{.9\textwidth}
%
%
\pgfplotsset{
  axis on top,
  y tick label style={rotate=0},
  legend style={inner xsep=1pt,inner ysep=0.5pt,nodes={inner sep=1pt,text depth=0.1em},font=\footnotesize},
  minor y tick num=0,
  minor x tick num=0,
 }
\centering

\newcommand\minus{%
  \scalebox{0.50}[1.0]{\( - \)}
}
\definecolor{mycolor1}{rgb}{0.00000,1.00000,1.00000}%
\begin{tikzpicture}

\begin{axis}[%
width=\figurewidth,
height=0.957\figureheight,
at={(0\figurewidth,0\figureheight)},
scale only axis,
xmin=-1,
xmax=1,
xtick={-1,-.8,-.6,-.4,-.2,0,.2,.4,.6,.8,1},
xlabel={Model parameters, $\theta = A$},
ymin=-250,
ymax=50,
ytick={-250,-200,-150,-100,-50,0,50},
axis background/.style={fill=white},
legend style={at={(0.02,0.98)},anchor=north west,legend cell align=left,align=left,draw=white!15!black}
]
\addplot [color=green,solid,line width=1.4pt]
  table[row sep=crcr]{%
-0.99	-212.65670863246\\
-0.97	-208.755975587387\\
-0.95	-204.860879719256\\
-0.93	-200.972106473577\\
-0.91	-197.09035311698\\
-0.89	-193.216328693775\\
-0.87	-189.350753971946\\
-0.85	-185.494361378182\\
-0.83	-181.647894921553\\
-0.81	-177.812110105448\\
-0.79	-173.987773827391\\
-0.77	-170.175664266332\\
-0.75	-166.376570757051\\
-0.73	-162.591293651272\\
-0.71	-158.82064416514\\
-0.69	-155.065444212662\\
-0.67	-151.326526224783\\
-0.65	-147.604732953719\\
-0.63	-143.900917262226\\
-0.61	-140.215941897463\\
-0.59	-136.550679249143\\
-0.57	-132.906011091661\\
-0.55	-129.282828309923\\
-0.53	-125.682030608603\\
-0.51	-122.104526204586\\
-0.49	-118.551231502363\\
-0.47	-115.02307075218\\
-0.45	-111.520975690757\\
-0.43	-108.045885164424\\
-0.41	-104.598744734549\\
-0.39	-101.180506265152\\
-0.37	-97.7921274926528\\
-0.35	-94.4345715776957\\
-0.33	-91.1088066390692\\
-0.31	-87.815805269741\\
-0.29	-84.5565440350841\\
-0.27	-81.3320029533982\\
-0.25	-78.1431649588748\\
-0.23	-74.9910153471936\\
-0.21	-71.8765412039773\\
-0.19	-68.800730816379\\
-0.17	-65.7645730681147\\
-0.15	-62.7690568183035\\
-0.13	-59.8151702645206\\
-0.11	-56.9039002905135\\
-0.0900000000000001	-54.036231799081\\
-0.0700000000000001	-51.2131470306601\\
-0.0499999999999999	-48.4356248682142\\
-0.03	-45.7046401290622\\
-0.01	-43.0211628443357\\
0.01	-40.386157526802\\
0.03	-37.8005824278313\\
0.05	-35.2653887843396\\
0.0700000000000001	-32.7815200565797\\
0.0900000000000001	-30.3499111576995\\
0.11	-27.9714876760309\\
0.13	-25.6471650911138\\
0.15	-23.3778479845025\\
0.17	-21.1644292464392\\
0.19	-19.0077892795197\\
0.21	-16.9087952005091\\
0.23	-14.8683000415014\\
0.25	-12.8871419516475\\
0.27	-10.9661434007031\\
0.29	-9.10611038567716\\
0.31	-7.30783164188156\\
0.33	-5.57207785970563\\
0.35	-3.8996009084538\\
0.37	-2.29113306860029\\
0.39	-0.747386273823857\\
0.41	0.730948635807025\\
0.43	2.14320264812524\\
0.45	3.4887292992594\\
0.47	4.76690537876394\\
0.49	5.97713161812426\\
0.51	7.11883336128967\\
0.53	8.19146121590267\\
0.55	9.19449168391402\\
0.57	10.127427770296\\
0.59	10.9897995685963\\
0.61	11.781164822106\\
0.63	12.5011094594518\\
0.65	13.1492481034624\\
0.67	13.7252245522036\\
0.69	14.228712231123\\
0.71	14.6594146152974\\
0.73	15.0170656208291\\
0.75	15.3014299644974\\
0.77	15.5123034908298\\
0.79	15.649513465823\\
0.81	15.7129188366097\\
0.83	15.7024104564371\\
0.85	15.6179112743926\\
0.87	15.459376489388\\
0.89	15.2267936679867\\
0.91	14.9201828257368\\
0.93	14.5395964717504\\
0.95	14.0851196163487\\
0.97	13.556869741672\\
0.99	12.9549967352367\\
};
\addlegendentry{$L_{\theta}(y_{1:T})$};

\addplot [color=red,dashed,line width=1.4pt]
  table[row sep=crcr]{%
-0.15	-254.101166980513\\
-0.13	-242.864936905449\\
-0.11	-231.665494333616\\
-0.0900000000000001	-220.499116354409\\
-0.0700000000000001	-209.363849561655\\
-0.0499999999999999	-198.259561311961\\
-0.03	-187.188023433577\\
-0.01	-176.153029315147\\
0.01	-165.160546186432\\
0.03	-154.218905340695\\
0.05	-143.339034074181\\
0.0700000000000001	-132.534734267689\\
0.0900000000000001	-121.823013848976\\
0.11	-111.224478899139\\
0.13	-100.763795956147\\
0.15	-90.4702361901044\\
0.17	-80.3783156575121\\
0.19	-70.5285488836607\\
0.21	-60.9683366943605\\
0.23	-51.7530136716762\\
0.25	-42.9470860330226\\
0.27	-34.6256973697647\\
0.29	-26.8763678380881\\
0.31	-19.8010624681476\\
0.33	-13.5186567637418\\
0.35	-8.167883382859\\
0.37	-3.9108633219922\\
0.39	-0.937349889313168\\
0.41	0.530154497721327\\
0.43	0.230207717316432\\
0.45	-2.145090784856\\
0.47	-6.95927162352677\\
0.49	-14.6427390476125\\
0.51	-25.7066337283077\\
0.53	-40.7600395138642\\
0.55	-60.5315093585702\\
0.57	-85.8962464766966\\
0.59	-117.910808214639\\
0.61	-157.858000978989\\
0.63	-207.30587017982\\
0.65	-268.186642051775\\
};
\addlegendentry{$Q_{\textup{ld}}(\theta,\theta_k)$};

\addplot [color=blue,dashdotted,line width=1.4pt]
  table[row sep=crcr]{%
-0.89	-251.229753838995\\
-0.87	-244.955514526374\\
-0.85	-238.756210784443\\
-0.83	-232.631842613201\\
-0.81	-226.582410012649\\
-0.79	-220.607912982785\\
-0.77	-214.70835152361\\
-0.75	-208.883725635125\\
-0.73	-203.134035317329\\
-0.71	-197.459280570222\\
-0.69	-191.859461393804\\
-0.67	-186.334577788075\\
-0.65	-180.884629753035\\
-0.63	-175.509617288684\\
-0.61	-170.209540395023\\
-0.59	-164.984399072051\\
-0.57	-159.834193319767\\
-0.55	-154.758923138173\\
-0.53	-149.758588527268\\
-0.51	-144.833189487052\\
-0.49	-139.982726017526\\
-0.47	-135.207198118688\\
-0.45	-130.50660579054\\
-0.43	-125.880949033081\\
-0.41	-121.33022784631\\
-0.39	-116.854442230229\\
-0.37	-112.453592184838\\
-0.35	-108.127677710135\\
-0.33	-103.876698806121\\
-0.31	-99.7006554727967\\
-0.29	-95.5995477101614\\
-0.27	-91.5733755182151\\
-0.25	-87.622138896958\\
-0.23	-83.7458378463901\\
-0.21	-79.9444723665112\\
-0.19	-76.2180424573216\\
-0.17	-72.5665481188211\\
-0.15	-68.9899893510097\\
-0.13	-65.4883661538874\\
-0.11	-62.0616785274542\\
-0.0900000000000001	-58.7099264717102\\
-0.0700000000000001	-55.4331099866554\\
-0.0499999999999999	-52.2312290722896\\
-0.03	-49.104283728613\\
-0.01	-46.0522739556255\\
0.01	-43.0751997533272\\
0.03	-40.1730611217179\\
0.05	-37.3458580607979\\
0.0700000000000001	-34.5935905705669\\
0.0900000000000001	-31.9162586510251\\
0.11	-29.3138623021725\\
0.13	-26.7864015240089\\
0.15	-24.3338763165345\\
0.17	-21.9562866797492\\
0.19	-19.6536326136531\\
0.21	-17.4259141182461\\
0.23	-15.2731311935282\\
0.25	-13.1952838394994\\
0.27	-11.1923720561598\\
0.29	-9.26439584350935\\
0.31	-7.41135520154802\\
0.33	-5.6332501302758\\
0.35	-3.93008062969275\\
0.37	-2.3018466997988\\
0.39	-0.74854834059397\\
0.41	0.729814447921697\\
0.43	2.13324166574826\\
0.45	3.46173331288564\\
0.47	4.71528938933392\\
0.49	5.89390989509306\\
0.51	6.99759483016308\\
0.53	8.02634419454392\\
0.55	8.98015798823569\\
0.57	9.85903621123829\\
0.59	10.6629788635518\\
0.61	11.3919859451761\\
0.63	12.0460574561113\\
0.65	12.6251933963574\\
0.67	13.1293937659144\\
0.69	13.5586585647821\\
0.71	13.9129877929608\\
0.73	14.1923814504503\\
0.75	14.3968395372508\\
0.77	14.526362053362\\
0.79	14.5809489987842\\
0.81	14.5606003735172\\
0.83	14.4653161775611\\
0.85	14.2950964109158\\
0.87	14.0499410735814\\
0.89	13.7298501655579\\
0.91	13.3348236868452\\
0.93	12.8648616374435\\
0.95	12.3199640173525\\
0.97	11.7001308265725\\
0.99	11.0053620651033\\
};
\addlegendentry{$Q_{\textup{ls}}(\theta,\theta_k)$};

\addplot [color=mycolor1,solid,line width=1.4pt]
  table[row sep=crcr]{%
-0.15	-263.809426774177\\
-0.13	-249.942854977411\\
-0.11	-236.440495162419\\
-0.0900000000000001	-223.259341141027\\
-0.0700000000000001	-210.363362925789\\
-0.0499999999999999	-197.723101318276\\
-0.03	-185.315076501564\\
-0.01	-173.121425028797\\
0.01	-161.129582773821\\
0.03	-149.332208855816\\
0.05	-137.726987772773\\
0.0700000000000001	-126.316620306017\\
0.0900000000000001	-115.109012985717\\
0.11	-104.117463356773\\
0.13	-93.3610484702702\\
0.15	-82.8651861476987\\
0.17	-72.6621505963787\\
0.19	-62.792129607941\\
0.21	-53.3040328189984\\
0.23	-44.2570701380283\\
0.25	-35.7220555284093\\
0.27	-27.7835967786244\\
0.29	-20.5424444205558\\
0.31	-14.1180763144534\\
0.33	-8.65188647172151\\
0.35	-4.31077672536631\\
0.37	-1.28985903037233\\
0.39	0.18102968477249\\
0.41	-0.181029684772493\\
0.43	-2.73461687370189\\
0.45	-7.91257683674334\\
0.47	-16.2220637027301\\
0.49	-28.2737081522276\\
0.51	-44.8161053287054\\
0.53	-66.7734043521912\\
0.55	-95.2927211418928\\
0.57	-131.807311562401\\
0.59	-178.121719631298\\
0.61	-236.527893297999\\
0.621839121291132	-280\\
};
\addlegendentry{$\bar{Q}(\eta)$};

\addplot [color=black,dashed]
  table[row sep=crcr]{%
0.7	-250\\
0.7	50\\
};
\addlegendentry{$\theta_{\textup{true}}$};

\addplot [color=black,solid]
  table[row sep=crcr]{%
0.4	-250\\
0.4	50\\
};
\addlegendentry{$\theta_k$};

\end{axis}
\end{tikzpicture}%
} \\
\caption{Lower bounds to the log likelihood $L_\theta(y_{1:T})$ of a first order system with a single unknown scalar parameter, $A$. $\qld\ththk$ and $\qls\ththk$ denote the bounds based on latent disturbances and states respectively, while $\bar{Q}(\eta)=Q_1(\tha,\theta_k)+Q_2(\thb,\theta_k)+\qbnd(\eta)$, where $\qbnd(\eta)$ is the bound based on Lagrangian relaxation defined in (\ref{eq:bound_on_Q3}).}
\label{fig:1D_bounds}
\end{figure}

In the following experiment, we investigate the fidelity of $Q(\theta,\theta_k)$ as a bound on $L_\theta(y_{1:T})$, as a function of the magnitude of the disturbances, $w_{1:T}$, and the choice of latent variables. As in Section \ref{sec:limit_cases}, we denote the bounds based on latent states and disturbances by $\qls\ththk$ and $\qld\ththk$, respectively. The results are presented in Figure \ref{fig:1D_bounds}, which depicts $\qls$, $\qld$ and $\loglik$ for a first order ($n_x=1$) LGSS model, each plotted as a function of the single unknown scalar parameter $\theta=A$.

\par We begin with the case of `small' disturbances (i.e. $\Sigma_w\ll\Sigma_v$) as depicted in Figure \ref{fig:1D_bounds}(a), and observe the following: $\qld\ththk$ represents $\loglik$ with high fidelity, whereas $\qls\ththk$ is localized about $\theta_k$. Such an observation is not without precedent. For instance, in the latent states formulation of \cite[Section 10]{Schon2011} it was noted that an initial disturbance covariance estimate $\Sigma_w=0$ results in $\theta_k=\theta_0$ for all $k$; i.e. the model parameters are not improved. This suggests that $\qls\ththk$ fails to accurately represent $\loglik$, except at $\theta=\theta_0$.  

\par Proposition \ref{prop:Qlvs_wSmall} makes this observation more precise: in the 1D case of Figure \ref{fig:1D_bounds}(a), when $\Sigma_w=0$, $\qls\ththk$ is undefined for $A\neq A_k$. Taken together, Figure \ref{fig:1D_bounds}(a) and Proposition \ref{prop:Qlvs_wSmall} suggest that as $\Sigma_w$ becomes smaller (relative to $\Sigma_v$) the bound $\qls\ththk$ becomes more localized about $\theta_k$, eventually collapsing to a single point when $\Sigma_w=0$. Conversely, as $\Sigma_w$ (and $\Sigma_1$) decrease, $\qld\ththk$ becomes an increasingly accurate representation of the log likelihood, eventually reproducing $\loglik$ \emph{exactly}, when $\Sigma_w$ (and $\Sigma_1$) are identically zero, as in Proposition \ref{prop:Qlvn_wSmall}.

\par Turning our attention to the case of `large' disturbances (i.e. $\Sigma_w\gg\Sigma_v$) as depicted in Figure \ref{fig:1D_bounds}(b), we observe the opposite behavior: $\qls\ththk$ faithfully represents the log likelihood, whereas $\qld\ththk$ appears to be localized about $\theta_k$. Once more, studying the limiting case $\Sigma_v=0$ offers insight into this behavior: Proposition \ref{prop:Qlvn_zSmall} states that when $\Sigma_v=0$, $\qld\ththk$ is undefined for $A\neq A_k$.

\par Taken together, Figure \ref{fig:1D_bounds}(b) and Proposition \ref{prop:Qlvn_zSmall} suggest that as $\Sigma_v$ decreases (i.e. as $\Sigma_w$ increases relative to $\Sigma_v$), the bound $\qld\ththk$ becomes more localized about $\theta_k$, eventually collapsing to a single point when $\Sigma_v=0$. Conversely, for this 1D experiment with $\theta=A$, Proposition \ref{prop:Qlvs_zSmall} states that $\qls\ththk$ will reproduce $L_\theta(y_{1:T})$ \emph{exactly}, when $\Sigma_v$ is identically zero. Indeed, in Figure \ref{fig:1D_bounds}(b) with $\Sigma_v\ll\Sigma_w$, we observe $\qls\ththk$  representing the likelihood faithfully.

\par To summarize: in the case of `large disturbances' (i.e. $\Sigma_w\gg\Sigma_v$), $\qld\ththk$ will tend to bound $L_\theta(y_{1:T})$ with greater fidelity, compared to $\qls\ththk$. In the case of `small disturbances' (i.e. $\Sigma_w\ll\Sigma_v$) the converse is true.

\subsection{Convergence rate}\label{sec:conv_rates}
It is clear from Figure \ref{fig:1D_bounds}(a), that both $\qld\ththk$ and $\qbnd(\eta)$ better represent $L_\theta(y_{1:T})$ compared to $\qls\ththk$. In fact, one would expect that optimization of $\qbnd(\eta)$, as in Algorithm \ref{alg:EMLVN}, would converge to $\theta^\text{ML}$ in fewer iterations than optimization of $\qls\ththk$, as in a `standard' EM algorithm \cite{gibson2005robust}. 

\par
\begin{figure}
\centering
\setlength{\figureheight}{.55\textwidth}\setlength{\figurewidth}{.9\textwidth}\input{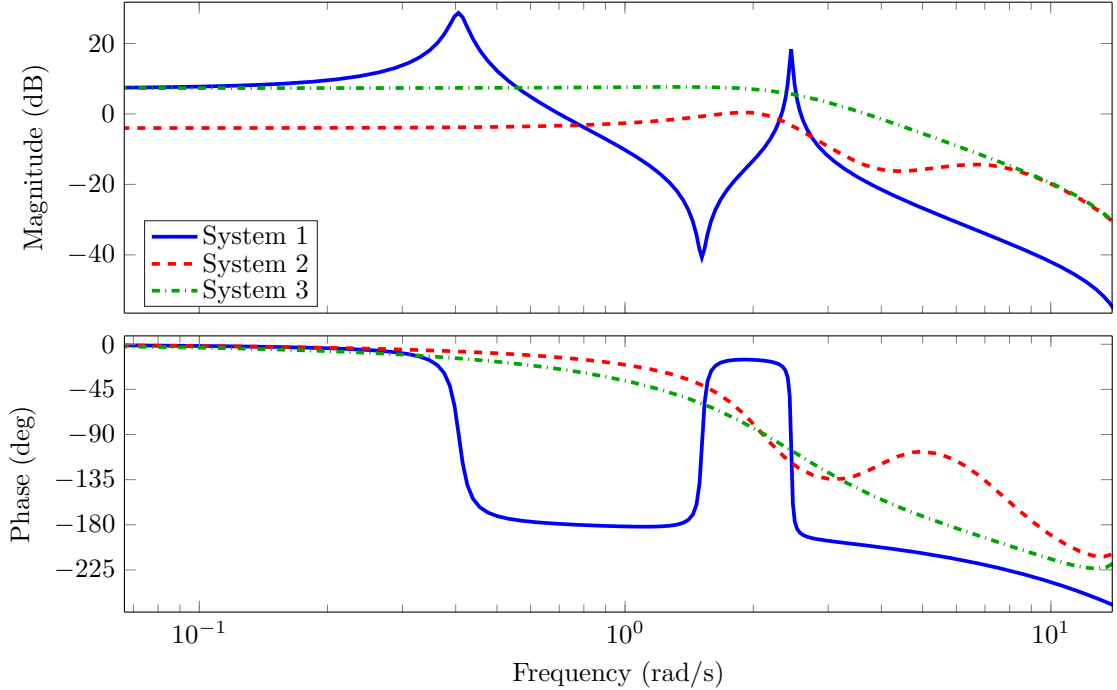}
\caption{Bode plots of $4\myth$ order systems used for the experiments presented in Figure \ref{fig:llc_new}.}
\label{fig:bodes}
\end{figure}

\begin{figure}[!ht]
  \centering
  \subfloat[System 1, sharp resonant peaks.]
    {\setlength{\figureheight}{.20\textheight}\setlength{\figurewidth}{.80\columnwidth}\input{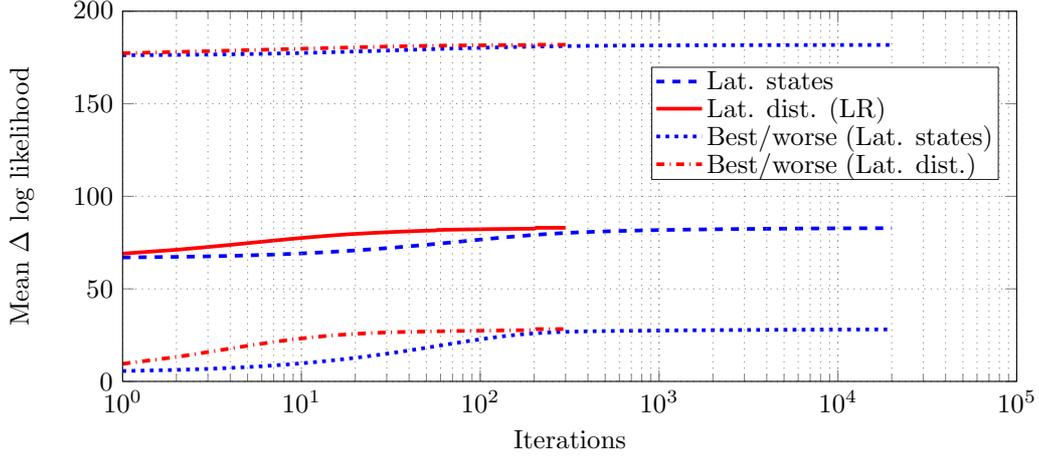}} \\
  \subfloat[System 2, smooth resonant peaks.]
    {\setlength{\figureheight}{.20\textheight}\setlength{\figurewidth}{.80\columnwidth}\input{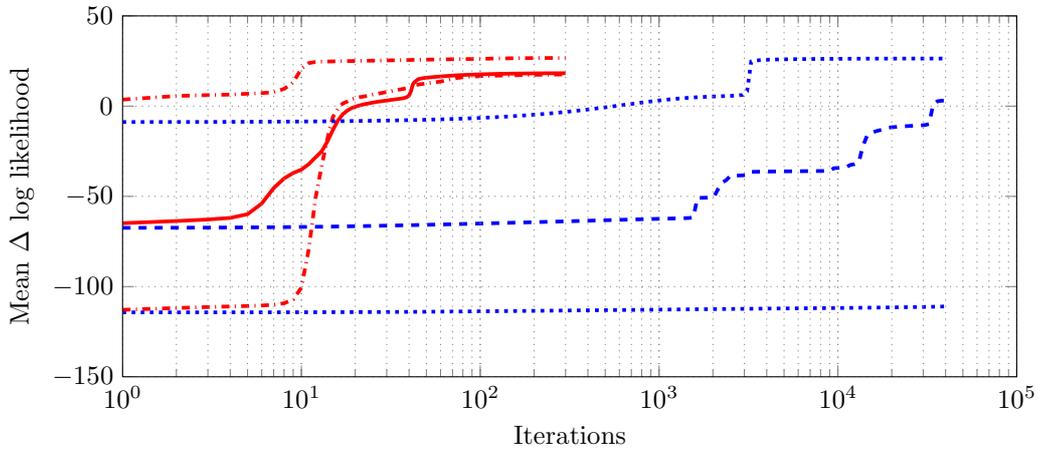}} \\
  \subfloat[System 3, overdamped.]
    {\setlength{\figureheight}{.20\textheight}\setlength{\figurewidth}{.80\columnwidth}\input{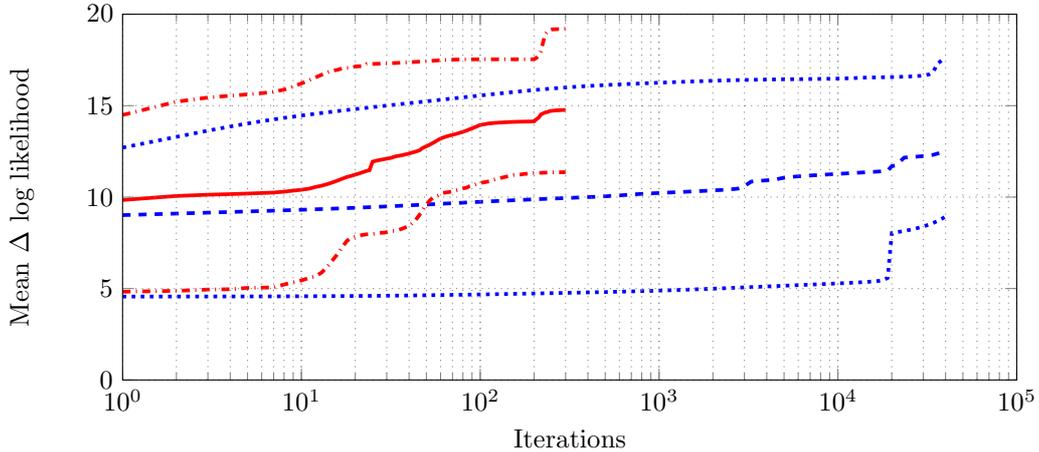}}
  \caption{Difference between $L_{\theta_k}(y_{1:T})$ and $L_{\theta_{\text{true}}}(y_{1:T})$ (where $\theta_{\text{true}}$ denotes the true model parameters) as a function of iterations for Algorithm \ref{alg:EMLVN} (EM with latent disturbances, red) and the method of \cite{gibson2005robust} (EM with latent states, blue dashed). The difference is averaged over 10 trials, each with SNR of 100, $\Sigma_v=1\times10^{-5}$ and $T=250$. Also plotted are the best and worst trial results (in terms of final likelihood) for each system. Bode plots for systems 1, 2 and 4 are depicted in Figure \ref{fig:bodes}.}
\label{fig:llc_new}
\end{figure}

 This principle, which is clearly understood in the first order example of Figure \ref{fig:1D_bounds}, is further illustrated in Figure \ref{fig:llc_new} for three different $4\myth$ order SISO systems; Bode plots for each system are given in Figure \ref{fig:bodes}. The results in Figure \ref{fig:llc_new} clearly show Algorithm \ref{alg:EMLVN}, based on latent disturbances, converging in fewer iterations than the latent states formulation of \cite{gibson2005robust}. These results are consistent with the analysis in Section \ref{sec:1dbounds}. Specifically, in each trial disturbances were `small' in magnitude ($\Sigma_w=1\times10^{-5}$) and so we expect $\qbnd(\eta)$ to better represent the likelihood, allowing Algorithm \ref{alg:EMLVN} to converge in fewer iterations.

\par It should be stressed that although Algorithm \ref{alg:EMLVN} converges in fewer iterations than the latent states formulation, each iteration is considerably more computationally expensive, and thus the total computation times for each algorithm are comparable. Nevertheless, this faster convergence rate is advantageous as it renders Algorithm 1 less sensitive to the choice of $\delta$ when termination conditions of the form (\ref{eq:term_heuristic}) are employed. Furthermore, as methods for SDP mature, one may expect Algorithm \ref{alg:EMLVN} to gain the upper hand in regards to computation time.    

\subsection{Stability of the identified model}
A desirable property of Algorithm \ref{alg:EMLVN} is that stability of the identified model is enforced at every iteration; recall from Lemma \ref{lem:stableModel} that we confine our search to an implicit parametrization of all stable models, which, by Lemma \ref{lem:finite_sup}, is necessary to ensure that $\lagj_\mult(\eta)$ is well-defined. Conversely, in a standard latent states implementation of the EM algorithm, the M step is accomplished by the solution of an unconstrained linear least squares problem \cite{gibson2005robust}. Consequently, it is possible that at any iteration (or indeed the conclusion) of the algorithm, the parameters $\theta_k$ could constitute an unstable model.

\par
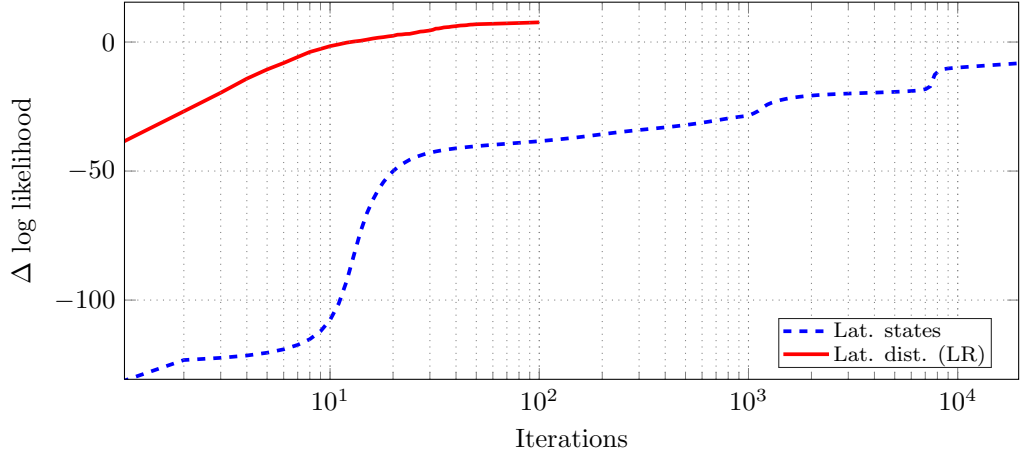
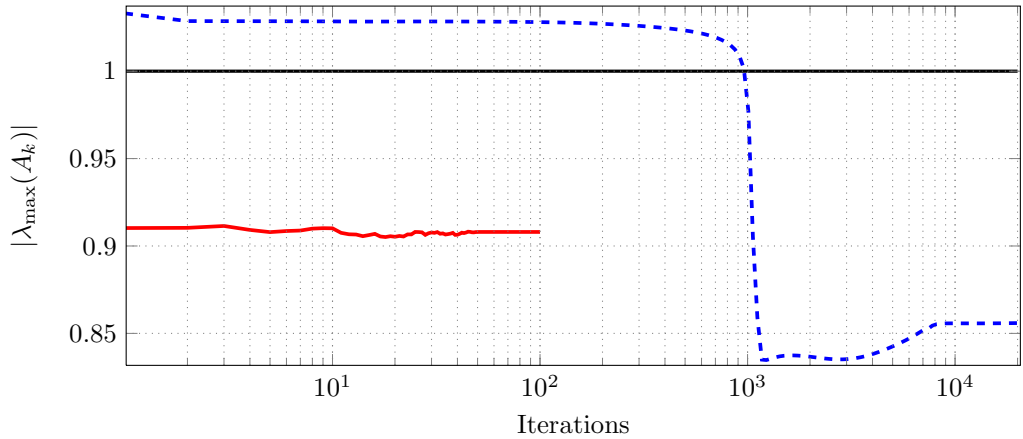
\begin{figure}[htb]
  \centering
  \subfloat[Difference between $L_{\theta_k}(y_{1:T})$ and $L_{\theta_{\text{true}}}(y_{1:T})$, where $\theta_{\text{true}}$ denotes the true model parameters, at each iteration.]
    {\setlength{\figureheight}{.35\columnwidth}\setlength{\figurewidth}{.80\textwidth}
%
%
\pgfplotsset{
  axis on top,
  y tick label style={rotate=0},
  legend style={inner xsep=1pt,inner ysep=0.5pt,nodes={inner sep=1pt,text depth=0.1em},font=\footnotesize},
  minor y tick num=0,
  minor x tick num=0,
 }
\centering

\newcommand\minus{%
  \scalebox{0.50}[1.0]{\( - \)}
}
\begin{tikzpicture}

\begin{axis}[%
width=\figurewidth,
height=0.972\figureheight,
at={(0\figurewidth,0\figureheight)},
scale only axis,
xmode=log,
xmin=1.04059349332624,
xmax=19564.7658374595,
xminorticks=true,
xlabel={Iterations},
xmajorgrids,
xminorgrids,
ymin=-130.711166594098,
ymax=15.4426795597478,
ytick={-100, -50, 0},
yticklabels={$-100$, $-50$, $0$},
ylabel={$\Delta$ log likelihood},
ymajorgrids,
axis background/.style={fill=white},
legend style={at={(0.97,0.03)},anchor=south east,legend cell align=left,align=left,draw=white!15!black}
]
\addplot [color=blue,dashed,line width=1.4pt]
  table[row sep=crcr]{%
1	-131.674750872395\\
2	-123.245209484045\\
3	-122.393425121958\\
4	-121.492345597388\\
5	-120.413117173462\\
6	-119.083180186319\\
7	-117.39143385704\\
8	-115.158070587745\\
9	-112.086877111817\\
10	-107.703641897188\\
11	-101.366339661712\\
12	-92.6780778585651\\
13	-82.56690707593\\
14	-73.351550674948\\
15	-66.3562484679301\\
16	-61.2034243408402\\
17	-57.2830177830597\\
18	-54.2545508037828\\
19	-51.899143667358\\
20	-50.0481896814353\\
21	-48.575476321973\\
22	-47.389698948115\\
24	-45.6308475440098\\
25	-44.9717029573668\\
26	-44.4189192510508\\
28	-43.5507067741051\\
29	-43.2054683458329\\
30	-42.9049112554674\\
32	-42.4073581547312\\
34	-42.0116479949221\\
35	-41.8422357864528\\
37	-41.5468536821655\\
39	-41.2967160823949\\
41	-41.080793672982\\
43	-40.8913124738311\\
46	-40.6447985431986\\
48	-40.4999361351361\\
50	-40.3673492099282\\
53	-40.1869876244112\\
56	-40.0242485024673\\
59	-39.8753005886113\\
62	-39.737324365833\\
65	-39.6081967757808\\
68	-39.4862869982266\\
72	-39.3328044502747\\
75	-39.2232241951081\\
79	-39.0830465136967\\
83	-38.9483610001393\\
88	-38.786067991245\\
92	-38.6600576933246\\
97	-38.5062602253161\\
102	-38.3556883349032\\
107	-38.2076044178216\\
113	-38.0324275395919\\
118	-37.8880999942684\\
124	-37.71651458552\\
131	-37.5182543252184\\
137	-37.3498877485798\\
144	-37.1554018992102\\
152	-36.936082507489\\
160	-36.7205756980389\\
168	-36.5097263647716\\
176	-36.3044458337087\\
185	-36.0812707264838\\
195	-35.8442504684823\\
205	-35.6198906138788\\
215	-35.4088928666463\\
226	-35.1924468487724\\
238	-34.9745478202015\\
250	-34.7744459495606\\
263	-34.5757166030755\\
276	-34.3932691488758\\
291	-34.1996043005192\\
305	-34.032024713115\\
321	-33.852647452379\\
337	-33.6831389547706\\
355	-33.5010850534615\\
373	-33.3255627051346\\
392	-33.1451789049244\\
412	-32.9589425248835\\
433	-32.7659876951182\\
455	-32.5656005373152\\
478	-32.3572306054478\\
503	-32.1314763452044\\
528	-31.9061893335871\\
555	-31.6633419639383\\
584	-31.4032615102952\\
613	-31.1445012929328\\
645	-30.8615605956371\\
678	-30.5743629226416\\
712	-30.2859790347254\\
749	-29.9849713128922\\
787	-29.6953446805872\\
827	-29.4183434061073\\
869	-29.1632695335204\\
914	-28.9289868362103\\
960	-28.7099909982533\\
1009	-28.3593894903823\\
1061	-27.6145040823133\\
1115	-26.7020968948837\\
1172	-25.2982424951806\\
1232	-24.2193374257914\\
1295	-23.4771238023506\\
1361	-22.9118406040302\\
1430	-22.4545666608069\\
1503	-22.0691817926223\\
1580	-21.7427943715205\\
1660	-21.4698655485862\\
1745	-21.2357096787076\\
1834	-21.0363697350863\\
1928	-20.8633657712015\\
2026	-20.713496148976\\
2130	-20.5801322894268\\
2238	-20.463131781063\\
2353	-20.3572405025892\\
2473	-20.2627877838053\\
2599	-20.1773343116533\\
2732	-20.0989117995635\\
2871	-20.0268289723732\\
3018	-19.9588827852867\\
3172	-19.8945106544911\\
3333	-19.8327083218897\\
3504	-19.7715656154076\\
3682	-19.7114959430667\\
3870	-19.6509054015507\\
4068	-19.5893455654331\\
4275	-19.5266857823346\\
4494	-19.4615775267674\\
4723	-19.3940953197608\\
4964	-19.3229346217\\
5217	-19.2470172844723\\
5483	-19.1642686522773\\
5763	-19.071208134432\\
6057	-18.9617130407149\\
6366	-18.8221359294966\\
6691	-18.6179744210493\\
7033	-18.2400195214917\\
7392	-17.1785442854447\\
7769	-12.661553781531\\
8165	-10.9784371509734\\
8582	-10.4822982582337\\
9020	-10.2381405510937\\
9480	-10.0646541230322\\
9964	-9.91707216503612\\
10472	-9.7812896028047\\
11007	-9.65113018830233\\
11568	-9.52451621753934\\
12158	-9.39970586299899\\
12779	-9.27588174918112\\
13431	-9.15297475267573\\
14116	-9.03072593238481\\
14837	-8.90887244876329\\
15594	-8.78775620449764\\
16389	-8.66742944185908\\
17226	-8.54771209290945\\
18105	-8.42904161612078\\
19029	-8.31143695783616\\
20000	-8.19507158472317\\
};
\addlegendentry{Lat. states};

\addplot [color=red,solid,line width=1.4pt]
  table[row sep=crcr]{%
1	-39.1151276955192\\
2	-26.8658163151415\\
3	-19.6631021783937\\
4	-14.1993575861559\\
5	-10.629112781605\\
6	-8.12372577060492\\
7	-5.81323287674888\\
8	-3.84997699251365\\
9	-2.6775452024347\\
10	-1.592602173351\\
11	-0.906797068876628\\
12	-0.236982302349219\\
13	0.196588575709626\\
14	0.503507128559576\\
15	0.928225481674701\\
16	1.35559038560359\\
17	1.67773489234767\\
18	1.90289731944644\\
19	2.1946351185147\\
20	2.40591801519852\\
21	2.84208760034818\\
22	2.92152023335508\\
23	3.11042139180933\\
24	3.1638314325691\\
25	3.45987867821304\\
26	3.73337666801409\\
27	4.04198914346851\\
28	4.13580867793671\\
29	4.31248322017066\\
30	4.48610328209614\\
31	4.73975768449849\\
32	5.17869428803651\\
33	5.23491751360365\\
34	5.41948003231236\\
35	5.65417138922602\\
36	5.74089726060535\\
37	5.81044662158321\\
38	5.97171174698263\\
39	6.07327939490953\\
40	6.15629286745181\\
41	6.29016743888472\\
42	6.39827976879624\\
43	6.43763958782117\\
44	6.47576116806145\\
45	6.54006071067154\\
46	6.67467711575959\\
47	6.75408524179824\\
48	6.77974199326604\\
49	6.84538144694672\\
51	6.92301894703557\\
52	6.94128405625573\\
53	6.95928623076527\\
54	6.977039321738\\
56	7.01184764326794\\
57	7.02892495010663\\
58	7.0457975463283\\
60	7.07896346796223\\
61	7.09527253228862\\
62	7.11140849580445\\
64	7.1431864982179\\
65	7.15884009698452\\
67	7.18970242592404\\
69	7.22000220610515\\
70	7.23495151589609\\
72	7.26446749139491\\
74	7.29349541241072\\
75	7.30783393732156\\
77	7.33617381734723\\
79	7.36408070951255\\
81	7.39157272773117\\
83	7.41866621529337\\
85	7.44537598554645\\
87	7.47171552412418\\
89	7.49769715965463\\
91	7.52333220849182\\
93	7.54863109793789\\
95	7.57360347157267\\
97	7.5982582796328\\
100	7.63466313692624\\
};
\addlegendentry{Lat. dist. (LR)};

\end{axis}
\end{tikzpicture}%
} \\
  \subfloat[Magnitude of the largest eigenvalue of $A_k$, at each iteration. When the spectral radius of $A_k$ is greater than unity, i.e. $|\lambda_\text{max}(A_k)|>1$, the model $\theta_k$ is unstable. ]
    {\setlength{\figureheight}{.35\columnwidth}\setlength{\figurewidth}{.80\textwidth}
%
%
\pgfplotsset{
  axis on top,
  y tick label style={rotate=0},
  legend style={inner xsep=1pt,inner ysep=0.5pt,nodes={inner sep=1pt,text depth=0.1em},font=\footnotesize},
  minor y tick num=0,
  minor x tick num=0,
 }
\centering

\newcommand\minus{%
  \scalebox{0.50}[1.0]{\( - \)}
}
\begin{tikzpicture}

\begin{axis}[%
width=\figurewidth,
height=0.926\figureheight,
at={(0\figurewidth,0\figureheight)},
scale only axis,
xmode=log,
xmin=1.01335209260209,
xmax=20630.8028553168,
xminorticks=true,
xlabel={Iterations},
xmajorgrids,
xminorgrids,
ymin=0.831705539358601,
ymax=1.03724489795918,
ytick={0.85, 0.9, 0.95, 1},
yticklabels={$0.85$, $0.9$, $0.95$, $\phantom{-00}1$},
ylabel={$|\lambda_{\textup{max}}(A_k)|$},
ymajorgrids,
axis background/.style={fill=white}
]
\addplot [color=black,solid,line width=1.2pt,forget plot]
  table[row sep=crcr]{%
1	1\\
2	1\\
3	1\\
4	1\\
5	1\\
6	1\\
7	1\\
8	1\\
9	1\\
10	1\\
11	1\\
12	1\\
13	1\\
14	1\\
15	1\\
16	1\\
17	1\\
18	1\\
19	1\\
20	1\\
21	1\\
22	1\\
24	1\\
25	1\\
26	1\\
28	1\\
29	1\\
30	1\\
32	1\\
34	1\\
35	1\\
37	1\\
39	1\\
41	1\\
43	1\\
46	1\\
48	1\\
50	1\\
53	1\\
56	1\\
59	1\\
62	1\\
65	1\\
68	1\\
72	1\\
75	1\\
79	1\\
83	1\\
88	1\\
92	1\\
97	1\\
102	1\\
107	1\\
113	1\\
118	1\\
124	1\\
131	1\\
137	1\\
144	1\\
152	1\\
160	1\\
168	1\\
176	1\\
185	1\\
195	1\\
205	1\\
215	1\\
226	1\\
238	1\\
250	1\\
263	1\\
276	1\\
291	1\\
305	1\\
321	1\\
337	1\\
355	1\\
373	1\\
392	1\\
412	1\\
433	1\\
455	1\\
478	1\\
503	1\\
528	1\\
555	1\\
584	1\\
613	1\\
645	1\\
678	1\\
712	1\\
749	1\\
787	1\\
827	1\\
869	1\\
914	1\\
960	1\\
1009	1\\
1061	1\\
1115	1\\
1172	1\\
1232	1\\
1295	1\\
1361	1\\
1430	1\\
1503	1\\
1580	1\\
1660	1\\
1745	1\\
1834	1\\
1928	1\\
2026	1\\
2130	1\\
2238	1\\
2353	1\\
2473	1\\
2599	1\\
2732	1\\
2871	1\\
3018	1\\
3172	1\\
3333	1\\
3504	1\\
3682	1\\
3870	1\\
4068	1\\
4275	1\\
4494	1\\
4723	1\\
4964	1\\
5217	1\\
5483	1\\
5763	1\\
6057	1\\
6366	1\\
6691	1\\
7033	1\\
7392	1\\
7769	1\\
8165	1\\
8582	1\\
9020	1\\
9480	1\\
9964	1\\
10472	1\\
11007	1\\
11568	1\\
12158	1\\
12779	1\\
13431	1\\
14116	1\\
14837	1\\
15594	1\\
16389	1\\
17226	1\\
18105	1\\
19029	1\\
20000	1\\
};
\addplot [color=blue,dashed,line width=1.4pt,forget plot]
  table[row sep=crcr]{%
1	1.03306830755947\\
2	1.02866077747764\\
3	1.02861760278393\\
4	1.02859906271489\\
5	1.02857997311384\\
6	1.02855991456764\\
7	1.02853862626795\\
8	1.02851584692998\\
9	1.02849134752486\\
10	1.02846507640957\\
11	1.02843753178042\\
12	1.02841050032507\\
13	1.02838780281329\\
14	1.02837422332219\\
15	1.02837159056087\\
16	1.02837694841761\\
17	1.02838579096756\\
18	1.02839499936714\\
19	1.02840319069609\\
20	1.02841004980563\\
21	1.02841568298739\\
22	1.02842027238369\\
24	1.02842687931733\\
25	1.02842909135663\\
26	1.02843067313166\\
28	1.0284321769981\\
29	1.02843219827198\\
30	1.02843178988211\\
32	1.02842983088236\\
34	1.02842655829105\\
35	1.0284244964925\\
37	1.02841963335399\\
39	1.02841390746813\\
41	1.02840744155927\\
43	1.02840033451974\\
46	1.0283886420764\\
48	1.02838024867126\\
50	1.02837143312206\\
53	1.02835750058779\\
56	1.02834280533766\\
59	1.02832742371218\\
62	1.02831141322432\\
65	1.02829481743384\\
68	1.0282776694272\\
72	1.02825398894818\\
75	1.02823563984093\\
79	1.02821041417915\\
83	1.02818433921159\\
88	1.02815057231316\\
92	1.02812263047346\\
97	1.02808655040135\\
102	1.0280491939185\\
107	1.02801056423612\\
113	1.02796253359152\\
118	1.02792112013944\\
124	1.02786977550634\\
131	1.02780763935121\\
137	1.02775251441897\\
144	1.02768610669364\\
152	1.02760759034047\\
160	1.02752647374059\\
168	1.02744298966663\\
176	1.02735739003208\\
185	1.02725888576118\\
195	1.02714712780134\\
205	1.02703339622301\\
215	1.02691811394881\\
226	1.02678994007588\\
238	1.02664891816601\\
250	1.02650700211826\\
263	1.02635252760182\\
276	1.02619744955851\\
291	1.02601781704638\\
305	1.02584941476516\\
321	1.02565585965732\\
337	1.0254608038108\\
355	1.02523905117787\\
373	1.02501420784823\\
392	1.02477273799279\\
412	1.02451303494847\\
433	1.02423312809926\\
455	1.0239306270356\\
478	1.0236026456574\\
503	1.0232304665587\\
528	1.02283951130923\\
555	1.02239296720371\\
584	1.021880726046\\
613	1.02132892577878\\
645	1.02066544117095\\
678	1.01990852575386\\
712	1.01903264322265\\
749	1.01793678043268\\
787	1.01660225652455\\
827	1.01486676702228\\
869	1.01246354132182\\
914	1.00864930655699\\
960	1.00142347711634\\
1009	0.977210792145781\\
1061	0.905376610715838\\
1115	0.859166873555277\\
1172	0.834977245503265\\
1232	0.834740676476857\\
1295	0.835510236229106\\
1361	0.836290966474773\\
1430	0.836863326500035\\
1503	0.837215598642512\\
1580	0.837378733887387\\
1660	0.837393583320437\\
1745	0.837295394969838\\
1834	0.837111370498337\\
1928	0.836861943794189\\
2026	0.836570810983123\\
2130	0.836253700892663\\
2238	0.835939634451426\\
2353	0.835644349641819\\
2473	0.835397490124047\\
2599	0.83521874262888\\
2732	0.835126530688951\\
2871	0.835136532862215\\
3018	0.835259519834979\\
3172	0.835500780865314\\
3333	0.835860034869041\\
3504	0.836341794089135\\
3682	0.836931367966358\\
3870	0.837629072841912\\
4068	0.838424745576597\\
4275	0.839302261277714\\
4494	0.840262393460286\\
4723	0.841285276888177\\
4964	0.842370338412713\\
5217	0.8435110300669\\
5483	0.844709415284179\\
5763	0.845973326518439\\
6057	0.847313972432484\\
6366	0.848756289633799\\
6691	0.850323007464404\\
7033	0.851899377468959\\
7392	0.852492022116453\\
7769	0.85441581266544\\
8165	0.855293057994121\\
8582	0.855610871097995\\
9020	0.8557122826135\\
9480	0.855743285205502\\
9964	0.855747203914517\\
10472	0.855741138410351\\
11007	0.855732773263449\\
11568	0.855725761152618\\
12158	0.85572180656945\\
12779	0.85572173427161\\
13431	0.855725894671395\\
14116	0.855734367401916\\
14837	0.855747101726777\\
15594	0.855763890055048\\
16389	0.855784475945193\\
17226	0.855808613664037\\
18105	0.855835909186728\\
19029	0.855866027170583\\
20000	0.855898589981862\\
};
\addplot [color=red,solid,line width=1.4pt,forget plot]
  table[row sep=crcr]{%
1	0.910260384717065\\
2	0.910373782583161\\
3	0.911443605414152\\
4	0.909186648682142\\
5	0.907930998854412\\
6	0.908560409809686\\
7	0.908833707090908\\
8	0.909923355237464\\
9	0.910188569944959\\
10	0.910107423881623\\
11	0.907512199091417\\
12	0.906693337845534\\
13	0.906551470467716\\
14	0.905603350006588\\
15	0.906258243978838\\
16	0.906938492125831\\
17	0.905434726565721\\
18	0.905112066179658\\
19	0.90558931433409\\
20	0.905264595887107\\
21	0.905732517441752\\
22	0.905416835360514\\
23	0.906625347769391\\
24	0.906625347769391\\
25	0.908024185977032\\
26	0.907971413859066\\
27	0.907801955710081\\
28	0.906342263305757\\
29	0.907265132011204\\
30	0.907722356493178\\
31	0.907493210647962\\
32	0.907949006426862\\
33	0.906987944579933\\
34	0.907206757885067\\
35	0.90656033366745\\
36	0.906783409765101\\
37	0.907062161447703\\
38	0.907491307491854\\
39	0.906343016841479\\
40	0.906424423377052\\
41	0.906959789385866\\
42	0.907615874588014\\
43	0.907373936811839\\
44	0.907690878880362\\
45	0.908177230051756\\
46	0.908070019313257\\
47	0.907825134108477\\
48	0.907782149912487\\
49	0.907960425077154\\
51	0.908022789923553\\
52	0.908022789923553\\
53	0.908022789923553\\
54	0.908022789923553\\
56	0.908022789923554\\
57	0.908022789923553\\
58	0.908022789923553\\
60	0.908022789923553\\
61	0.908022789923554\\
62	0.908022789923553\\
64	0.908022789923552\\
65	0.908022789923554\\
67	0.908022789923553\\
69	0.908022789923553\\
70	0.908022789923553\\
72	0.908022789923555\\
74	0.908022789923552\\
75	0.908022789923553\\
77	0.908022789923553\\
79	0.908022789923554\\
81	0.908022789923553\\
83	0.908022789923554\\
85	0.908022789923553\\
87	0.908022789923553\\
89	0.908022789923554\\
91	0.908022789923553\\
93	0.908022789923554\\
95	0.908022789923554\\
97	0.908022789923554\\
100	0.908022789923554\\
};
\end{axis}
\end{tikzpicture}%
}
  \caption{Log likelihood and spectral radius of $A_k$ at each iteration for two different EM algorithms: i. EM with latent states as in \cite{gibson2005robust} (Lat. states, blue dash); ii. Algorithm \ref{alg:EMLVN} (Lat. dist. (LR), red).	The spectral radius of $A$ for the true system was 0.90.}
  \label{fig:maxEig2}
\end{figure}

 Such a scenario is illustrated in the numerical experiment of Figure \ref{fig:maxEig2}, which depicts the identification of a $4\myth$ order model, similar to System 2 in Figure \ref{fig:bodes}. 
From Figure \ref{fig:maxEig2}(b) it is apparent that, for the first one thousand iterations, the parameters maintained by the latent states EM algorithm represent an \emph{unstable} model (i.e. $|\mult_\text{max}(A_k)|>1$). This instability can be particularly problematic, given the slow convergence rate; e.g. in this instance, if a heuristic such as (\ref{eq:term_heuristic}) was used employed, for $\delta>4.7\times10^{-3}$ the algorithm would terminate before the thousandth iteration, and an unstable model would be returned. Conversely, the parameters maintained by Algorithm \ref{alg:EMLVN} constitute a stable model at each iteration. 

\section{Conclusion}
In this paper, we have formulated the EM algorithm over latent disturbances, rather than states, for the identification of linear dynamical systems. Our main contribution is the use of Lagrangian relaxation to obtain a convex approximation of the challenging maximization step, guaranteed not to decrease the likelihood at each iteration. Though more computationally complex, this formulation with latent disturbances allows EM to be applied to singular state-space models, where latent states based methods break down. 

\par Extension of this approach to the identification of nonlinear models shall be the subject of future research. In the nonlinear case, two major challenges arise during the formulation of EM with latent disturbances. First, the E step (c.f. Section \ref{sec:EM_estep}) now involves a nonlinear disturbance smoothing problem, for which no closed form solution is known to exist. In recent decades, \emph{sequential Monte Carlo} (SMC) methods \cite{gordon1993novel} have emerged as effective tools for overcoming similar difficulties, having already proved useful in nonlinear, non-Gaussian \emph{state smoothing} \cite{schon2015sequential} and \emph{disturbance filtering} \cite{murray2013disturbance} problems. 

\par Second, nonlinearity of the model complicates the Lagrangian relaxation of the M step; e.g. the bound $\hat{J}_\mult(\eta)$ cannot be evaluated analytically, as the supremum (in (\ref{eq:Jsup})) requires optimization of a function that is no longer quadratic in $x$. To proceed, one might approximate the simulation error terms in $Q_3(\thc,\theta_k)$ with the \emph{linearized simulation error}, introduced in \cite{Tobenkin2010}, to which the Lagrangian relaxation presented in this work can be applied with little modification. Alternatively, when the system nonlinearity is modeled as a polynomial, \emph{sum-of-squares} (SOS) programing \cite{Parrilo2000} may be used to generate, and optimize, convex approximations to the Lagrangian relaxation.    

\bibliographystyle{IEEETran}
\bibliography{sysid_2}

\end{document}